\documentclass[twocolumn,notitlepage,prl,superscriptaddress,longtable,longbibliography]{revtex4-2}
\usepackage{mhchem}
\usepackage{amsthm}
\usepackage{amsmath}
\usepackage{amssymb}
\usepackage{mathdots}
\usepackage{graphicx}
\usepackage{mathrsfs}
\usepackage{longtable}
\usepackage{multirow}
\usepackage{babel}
\usepackage{amsthm} 
\usepackage{xcolor}
\usepackage{bm}
\usepackage{hyperref}
 
\makeatletter

\newcommand{\Sym}{\operatorname{Sym}}

\newcommand{\Id}{\mathbb I}

\theoremstyle{definition}

\begin{document}

\title{Moment-Selective Quantum Designs in Aperiodic Temporal Ensembles}

\author{Yang Peng}\email{yang.peng@csun.edu}
\affiliation{Department of Physics and Astronomy, California State University, Northridge, Northridge, California 91330, USA}
\affiliation{Institute of Quantum Information and Matter and Department of Physics, California Institute of Technology, Pasadena, CA 91125, USA}

\begin{abstract}
We introduce a general mechanism for engineering moment-selective temporal
designs with hierarchically structured aperiodic drives.  At special control
settings where the driven dynamics is constrained by a finite symmetry group,
lower-order moments can already reproduce Haar statistics while a selected
higher-order deviation remains nonzero.  A small static detuning from such a
setting generically produces an exponentially long lifetime whose exponent
scales inversely with the square of the detuning.  We illustrate this mechanism
in qubit systems through exact Fibonacci and silver-mean constructions with
tetrahedral and icosahedral resonances, for which the first retained non-Haar
moments occur at orders $3$ and $6$, respectively.  We further show that
stochastic control fluctuations produce a qualitatively different cutoff from
coherent detuning.  Finally, we propose a finite-time signature of the
long-lived moment that can be measured without requiring observations
over exponentially long times.
\end{abstract}

\maketitle

\emph{Introduction.---}
Long-lived modes are a basic resource in nonequilibrium physics: they determine
how long information survives and often signal conservation laws, localization,
kinetic bottlenecks, or prethermal regimes
~\cite{DAlessio2014,Lazarides2014,Ponte2015,Abanin2015,Mori2016,Abanin2017,Else2017}.
Most familiar mechanisms retain information associated with a conserved
quantity, an emergent slow degree of freedom, or an approximate symmetry.
Here we ask a different control question: can a drive rapidly erase coarse
statistical information while deliberately retaining a selected finer feature
for a much longer time?

Quantum designs provide a natural language for this question.  A pure-state
$k$-design reproduces the first $k$ moments of the uniform (Haar) distribution
of pure states~\cite{Scott2006,AmbainisEmerson2007,RoyScott2007}; higher moments
therefore probe progressively finer statistical structure.  Hilbert-space
ergodicity (HSE) turns this hierarchy into a property of a single driven
trajectory~\cite{Pilatowsky2024}: the states visited successively by the
system, sampled uniformly in time, form a \emph{temporal ensemble}, and
$k$-HSE means that this ensemble reproduces Haar statistics through moment
order $k$.  Fibonacci driving can realize complete HSE, in which every
finite-order moment approaches its Haar value asymptotically
~\cite{Pilatowsky2023}.  Different moments, however, need not approach Haar
statistics on the same time scale~\cite{Pilatowsky2025,Liu2026}, and recent
NV-center experiments show that such moment-resolved dynamics can be measured
directly~\cite{Pan2026}.  Related work connects temporal ensembles to deep
thermalization, noise learning, subspace ergodicity, and temporal designs
~\cite{Mark2024,Shaw2025,Ghosh2025,Logaric2025,Zhou2026}.  Here we ask whether
a temporal ensemble can realize a \emph{moment-selective temporal design}, in
which lower moments become Haar-like while one chosen higher-order deviation
remains detectable for a parametrically longer time.

\begin{figure*}[t]
 \includegraphics[width=\textwidth]{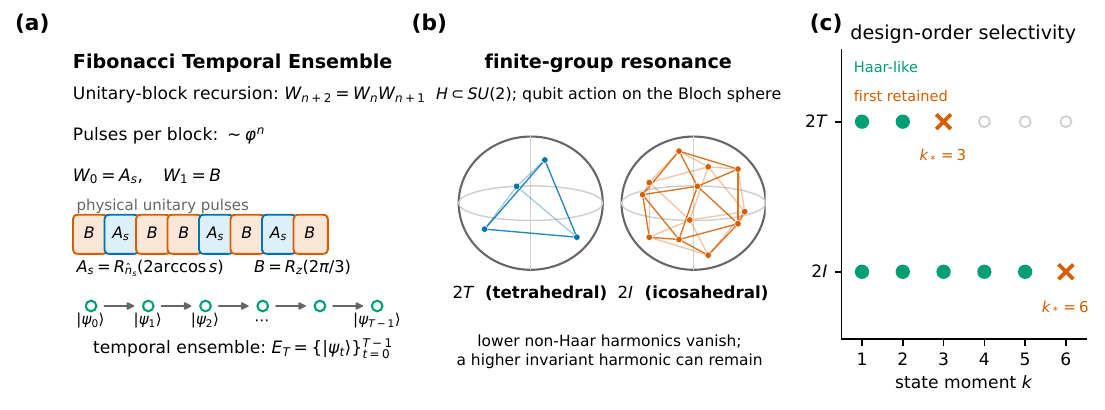}
 \caption{Moment-selective temporal design from a Fibonacci drive.
 (a) If $W_n$ denotes the total unitary generated by the $n$th Fibonacci
 pulse block, then $W_0=A_s$, $W_1=B$, and $W_{n+2}=W_nW_{n+1}$; the
 successive states generated by the physical pulse sequence form the temporal
 ensemble $\mathcal E_T$.
 (b) At finite-group resonances, the pulse unitaries generate the binary
 tetrahedral ($2T$) or binary icosahedral ($2I$) group.  Their action on the
 Bloch sphere eliminates lower non-Haar harmonics while a higher invariant
 harmonic can remain.
 (c) Design-order selectivity: the $2T$ resonance is Haar-like through state
 moment $k=2$ and first retains a non-Haar component at $k_*=3$, whereas the
 $2I$ resonance is Haar-like through $k=5$ and first retains one at $k_*=6$.}
 \label{fig:overview}
\end{figure*}

Aperiodic driving provides a natural route to this separation of time scales.
Consider a qubit driven by two elementary pulse unitaries, $A$ and $B$.  Let
$W_n$ denote the total unitary generated by the $n$th Fibonacci pulse block.
With $W_0=A$ and $W_1=B$, chronological multiplication gives
$W_{n+2}=W_nW_{n+1}$, where later pulses multiply on the left.  The
corresponding pulse block contains a Fibonacci number of physical pulses and
therefore grows exponentially with $n$ [Fig.~\ref{fig:overview}(a)].  A
statistical feature that changes only weakly from one block generation to the
next can consequently survive for an exceptionally large number of applied
pulses.  Structured and quasiperiodic
drives are already known to exhibit slow heating and unusual relaxation
~\cite{Dumitrescu2018,Else2020,Crowley2019,Zhao2021,Lapierre2020,Romanelli2009,Mo2025,LiuNature2026}.
Fibonacci driving, in particular, can 
produce unconventional steady states and self-similar dynamical structures
that reflect the recursive organization of the
drive~\cite{Nandy2018,Maity2019,Schmid2025};
our goal here is to make this hierarchical dynamics selective in moment order.

The selectivity arises at special pulse settings where compositions of the
qubit rotations generate a finite subgroup of $SU(2)$ rather than exploring
the full group.  We call such a setting a \emph{finite-group resonance}.  At
an appropriate resonance, the finite symmetry is sufficient to reproduce the
Haar values of the lower moments while allowing a selected higher-order
deviation to remain nonzero
~\cite{Delsarte1977,Scott2006,RoyScott2007,Reznick1995}.  A generic small
static detuning from resonance produces only a quadratic loss of this
higher-order component per block via the recursion relation.  The finite-group
mechanism and its moment-order selectivity are summarized in
Fig.~\ref{fig:overview}(b,c).  Because the block lengths grow exponentially,
the exponent of the resulting lifetime grows inversely with the square of
the detuning.  Symmetry or fine tuning can suppress the leading loss still
further.

We establish the recursive averaging framework for a broad class of recursively generated
aperiodic drives and give exact Fibonacci and silver-mean qubit realizations
with tetrahedral and icosahedral finite-group resonances.  We then propagate
the actual aperiodic pulse sequences to demonstrate the separation directly
in physical time.  Finally, we distinguish a static detuning from
pulse-to-pulse fluctuations of the same experimental control parameter:
independent fluctuations give a noise-limited lifetime proportional to the
inverse noise variance, rather than the exponentially long static-detuning
lifetime.  We also propose a direct implementation using a single
nitrogen-vacancy (NV) center in diamond and a finite-time spectroscopic test
based on the narrowing range of pulse settings for which the selected
higher-order signal remains detectable.

\emph{Moment-selective temporal designs.---}
After $T$ pulses, the trajectory defines the temporal ensemble
$\mathcal E_T=\{|\psi_t\rangle\}_{t=0}^{T-1}$, where $|\psi_t\rangle$ is the
state after pulse $t$.  Its $k$th moment and the corresponding Haar moment are
\begin{equation}
 \begin{aligned}
 M_k(T)&=\frac{1}{T}\sum_{t=0}^{T-1}
 (|\psi_t\rangle\langle\psi_t|)^{\otimes k},\\
 M_k^{\rm Haar}&=\int d\psi_{\rm Haar}
 (|\psi\rangle\langle\psi|)^{\otimes k}.
 \end{aligned}
 \label{eq:state_moment}
\end{equation}
An exact state $k$-design satisfies $M_j=M_j^{\rm Haar}$ for every $j\le k$.
At finite observation time, we instead seek a \emph{moment-selective temporal
design}: the lower moments are already close to their Haar values while one
selected higher moment remains non-Haar.  We denote by $k_*$ the first moment
order allowed to retain such a deviation.  The regime of interest is therefore
$M_j(T)\simeq M_j^{\rm Haar}$ for $j<k_*$, while the $k_*$th moment remains
appreciably different from $M_{k_*}^{\rm Haar}$.

We focus on qubits in the main text; the corresponding formulation for more
general quantum systems is given in the Supplemental Material~\cite{SM}.
For a qubit in a pure state, each state corresponds to a Bloch vector
$\bm n$ on the unit sphere.  The angular structure of the temporal ensemble
can be resolved using spherical harmonics $Y_{\ell m}(\bm n)$, where $\ell$
is the harmonic degree and $m=-\ell,\ldots,\ell$.  The degree $\ell$ sets the
angular resolution: $\ell=1$ describes dipolar structure, $\ell=2$
quadrupolar structure, and larger $\ell$ probe progressively finer angular
features~\cite{Delsarte1977,Bannai2015}.  We quantify the total deviation at
degree $\ell$ by the normalized rotationally invariant amplitude
\begin{equation}
 R_\ell(T)=
 \sqrt{\frac{4\pi}{2\ell+1}}
 \left[\sum_{m=-\ell}^{\ell}
 \left|\frac1T\sum_{t=0}^{T-1}Y_{\ell m}(\bm n_t)\right|^2\right]^{1/2}.
 \label{eq:observable}
\end{equation}
The normalization gives $0\le R_\ell(T)\le1$. 
For the Haar-uniform distribution, the average of every nonconstant spherical
harmonic vanishes,
$\int d\bm n_{\rm Haar}\,Y_{\ell m}(\bm n)=0$ for $\ell>0$.
Accordingly, a trajectory whose temporal distribution approaches the Haar
distribution satisfies $R_\ell(T)\to0$ as $T\to\infty$.  At finite $T$,
however, even $T$ independent Haar-random states give a nonzero sampling
fluctuation.  With the normalization in Eq.~\eqref{eq:observable},
$ \left\langle R_\ell^2(T)\right\rangle_{\rm Haar} =1/T$,  $\ell>0$.
Thus $R_\ell(T)$ should be compared with the finite-sampling Haar scale
$T^{-1/2}$ when deciding whether a measured harmonic component remains
detectably non-Haar. 

At a finite-group resonance, let $H$ denote the finite rotation group
generated by the pulse unitaries on the Bloch sphere.  Averaging over the
states related by $H$ eliminates the harmonic components below a first degree
$\ell_*$, while at degree $\ell_*$ there is a nonzero combination of
spherical harmonics that is unchanged by every rotation in $H$.  We define
$\ell_*$ as the smallest degree with such a surviving component.  Consequently,
the $R_\ell$ with lower $\ell$ can vanish while $R_{\ell_*}$ remains nonzero.  In the
qubit realizations below, $\ell_*$ coincides with the first non-Haar moment
order, $\ell_*=k_*$.  We use $R_{\ell_*}(T)$ as the observable measure of the
selected higher-order statistical memory.

To move away from resonance, we consider a smooth one-parameter family of
pulse settings that passes through the resonant point.  We denote by $\delta$
the displacement of this control parameter from its resonant value, with
$\delta=0$ at resonance.  Away from resonance, the finite-group symmetry is
no longer exact, and the surviving harmonic component decays under successive
block generations.  We characterize this block-level decay by an exponent
$\Delta_{\ell_*}$: after $n$ block generations, the slowly decaying
$\ell_*$ component is reduced by a factor $e^{-n\Delta_{\ell_*}}$.  Thus
$\Delta_{\ell_*}=0$ at exact resonance, whereas
$\Delta_{\ell_*}>0$ describes loss of the selected higher-order memory.

For a generic small detuning along a smooth family for which the
finite-dimensional block recursion returns after a fixed number of generations,
the decay exponent turns on quadratically,
\begin{equation}
 \Delta_{\ell_*}(\delta)
 =
 c\,\delta^2+O(\delta^3),
 \qquad c>0.
 \label{eq:quad}
\end{equation}
Physically, a small detuning $\delta$ mixes this
component with other harmonic components by an amount linear in $\delta$.
Because the resulting loss is set by the squared mixing amplitude, the decay
rate begins at order $\delta^2$.
If symmetry or fine tuning suppresses this leading amplitude so that it first
appears at order $\delta^q$, the decay instead begins as
$\Delta_{\ell_*}\propto\delta^{2q}$.  The precise conditions and general
proof are given in the Supplemental Material~\cite{SM}.

The recursive generation of pulse blocks converts this weak decay per generation into a much
longer lifetime measured in physical pulses.  The number of pulses in
the $n$th block generation grows exponentially as
$\propto\lambda^n$, with $\lambda>1$ fixed by the aperiodic sequence
construction~\cite{DumontThomas1989,BaakeGrimmJoseph1993}; for Fibonacci,
$\lambda=\varphi=(1+\sqrt5)/2$.  Since
$\Delta_{\ell_*}\propto\delta^2$, an order-one reduction of the selected
component requires a number of block generations proportional to
$\delta^{-2}$, which corresponds to exponentially many physical pulses.

For a fixed small threshold $\epsilon$, we define
$\tau_{k_*,\epsilon}$ as the pulse number beyond which
$R_{\ell_*}(T)$ remains below $\epsilon$ at all later times.  Because
$\ell_*=k_*$ in the qubit examples considered here, this is the lifetime of
the selected $k_*$th-order deviation.  Combining the quadratic block-level
decay with exponential block growth gives
$\log\tau_{k_*,\epsilon}\propto\delta^{-2}$ for sufficiently small $|\delta|$.
More generally,
$\Delta_{\ell_*}\propto\delta^{2q}$ gives
$\log\tau_{k_*,\epsilon}\propto\delta^{-2q}$.  

\emph{Exact analytical realizations.---}
Consider the Fibonacci drive generated by the two pulse unitaries
\begin{equation}
 A_s=R_{\hat{\bm n}_s}(2\arccos s), \quad B=R_z(2\pi/3),
\end{equation}
where $R_{\hat{\bm n}}(\theta)$ denotes a rotation by angle $\theta$ about
the Bloch-sphere axis $\hat{\bm n}$, $s$ is a tunable control parameter, and
\begin{equation}
 \hat{\bm n}_s=\frac{1}{\sqrt{1-s^2}}
 \left(
 \sqrt{\frac{2(1+s)(1-2s)}{3}},
 0,
 -\frac{1+s}{\sqrt{3}}
 \right).
\end{equation}
Varying $s$ changes both the rotation angle and the rotation axis of $A_s$,
thereby defining a smooth one-parameter family of Fibonacci drives.  
This family contains four finite-group resonances,
\begin{equation}
 s_*\in\{-\varphi/2,-1/2,0,1/(2\varphi)\}.
 \label{eq:points}
\end{equation}
The points $s_*=-1/2$ and $s_*=0$ generate the binary tetrahedral group $2T$,
the double cover in $SU(2)$ of the tetrahedral rotation group, whereas
$s_*=-\varphi/2$ and $s_*=1/(2\varphi)$ generate the binary icosahedral group
$2I$, the corresponding double cover of the icosahedral rotation group~\cite{SM}.

\begin{figure*}[t]
 \includegraphics[width=\textwidth]{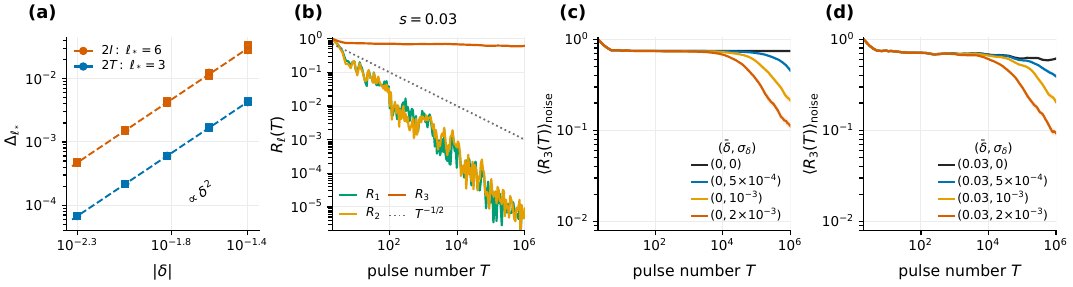}
 \caption{Detuning and noise dependence of the selected harmonic in the
 Fibonacci drive.
 (a) Block-level decay exponent $\Delta_{\ell_*}$ versus static detuning
 $|\delta|$ near the binary tetrahedral ($2T$, $\ell_*=3$) and binary
 icosahedral ($2I$, $\ell_*=6$) resonances.  The dashed lines show the
 quadratic scaling $\Delta_{\ell_*}\propto\delta^2$.
 (b) Direct propagation of the actual nonrepeated Fibonacci sequence at
 $s=0.03$: $R_1(T)$, $R_2(T)$, and $R_3(T)$ are shown together with the
 finite-sampling Haar scale $T^{-1/2}$.  The selected $R_3$ remains order one
 through $10^6$ pulses while the lower harmonics become small.
 (c,d) Noise-averaged selected signal $\langle R_3(T)\rangle_{\rm noise}$ for
 independent pulse-to-pulse fluctuations with mean detuning $\bar\delta=0$
 in (c) and $\bar\delta=0.03$ in (d), for the indicated values of
 $\sigma_\delta$.  The $\sigma_\delta=0$ curves give the corresponding
 deterministic limits.}
 \label{fig:results}
\end{figure*}

At the tetrahedral resonances, the first surviving non-Haar harmonic occurs
at $\ell_*=k_*=3$, whereas at the icosahedral resonances it occurs at
$\ell_*=k_*=6$~\cite{Delsarte1977,Reznick1995}.  These resonances therefore
produce approximate state-2- and state-5-design windows, respectively
[Fig.~\ref{fig:overview}(c)].  For a resonance at $s=s_*$, we define
$\delta=s-s_*$.  Expanding the exact Fibonacci block evolution about each
resonance verifies Eq.~\eqref{eq:quad}, with $c>0$ determined by the
corresponding finite-dimensional recursion [Fig.~\ref{fig:results}(a)].

The construction extends beyond Fibonacci driving.  As an independent
example, consider the silver-mean recursion for the total unitary blocks,
$W_{n+1}=W_{n-1}W_n^2$~\cite{KolarAli1990,BaakeGrimmJoseph1993,
Romanelli2009,Lapierre2020}.  Its icosahedral resonance uses the same pulse
pair as the Fibonacci resonance at $s=1/(2\varphi)$, but away from resonance
the pulses follow a different one-parameter family adapted to the
silver-mean recursion.  A local detuning along this family also obeys
Eq.~\eqref{eq:quad} for $\ell_*=6$, with a positive coefficient determined
by the silver-mean block recursion.

The Supplemental Material~\cite{SM} provides more technical details 
of the results on the Fibonacci and silver-mean drives.

\emph{Direct dynamics and imperfections.---}
We now test the separation directly in physical time by propagating the actual
nonrepeated Fibonacci pulse sequence, without using the block recursion to generate the
dynamics.  For the numerical results below we take the initial qubit state to
be $|\psi_0\rangle=|0\rangle$, corresponding to the north-pole Bloch vector
$\bm n_0=\hat{\bm z}$.

Near the tetrahedral Fibonacci resonance at $s_*=0$, choosing $s=0.03$ gives
after $T=10^6$ pulses
$R_3=0.615$, while $R_1=5.9\times10^{-6}$ and
$R_2=8.9\times10^{-6}$ [Fig.~\ref{fig:results}(b)].  For comparison, the
same initial state gives $R_3=0.745$ at the exact resonance.  Thus the
third-order non-Haar signal remains pronounced after one million pulses even
though the lower angular moments are already negligible.  Figure~\ref{fig:results}
focuses on the direct Fibonacci dynamics; the independent silver-mean
realization and its $\ell_*=6$ detuning law are derived in the Supplemental
Material~\cite{SM}.

We next model fluctuations of the same experimental control parameter used
to tune the resonance.  Let $\bar\delta$ denote the mean, time-independent
detuning and let $\epsilon_t$ denote an independent fluctuation on the $t$th
application of the tunable pulse.  We take
$\delta_t=\bar\delta+\epsilon_t$, with
$\langle\epsilon_t\rangle=0$ and
$\langle\epsilon_t\epsilon_{t'}\rangle
=\sigma_\delta^2\delta_{tt'}$, so $\sigma_\delta$ is the root-mean-square
pulse-to-pulse fluctuation of the detuning parameter.  For the Fibonacci
family above, this means that each application of $A_s$ uses
$s_t=s_*+\bar\delta+\epsilon_t$, while $B$ is kept fixed.  Because
$\delta=s-s_*$ for this family, both $\bar\delta$ and $\sigma_\delta$ are
dimensionless control-parameter displacements.

At exact mean resonance, $\bar\delta=0$, these independent fluctuations give
the otherwise persistent $R_{\ell_*}$ signal a finite lifetime.  Averaging
$R_{\ell_*}(T)$ over independent noisy trajectories, the decay becomes
faster as $\sigma_\delta$ increases [Fig.~\ref{fig:results}(c)], consistent
with the weak-noise scale
$\tau_{\rm noise}\sim\sigma_\delta^{-2}$.  For example, at the tetrahedral
resonance with $\sigma_\delta=10^{-3}$, averaging over 64 independent noise
realizations gives
$\langle R_3(10^6)\rangle_{\rm noise}=0.213\pm0.018$, whereas
$\langle R_1(10^6)\rangle_{\rm noise}=5.8\times10^{-6}$ and
$\langle R_2(10^6)\rangle_{\rm noise}=7.9\times10^{-6}$.  Thus stochastic
control errors shorten the selected higher-order memory while leaving the
already-small lower-order signals essentially unchanged.

The stochastic-noise lifetime is qualitatively different from the lifetime
generated by a static detuning.  When $\sigma_\delta=0$ and
$\bar\delta\ne0$, the deterministic hierarchy gives
$\log\tau_{k_*,\epsilon}\sim\bar\delta^{-2}$; when
$\bar\delta=0$ and $\sigma_\delta\ne0$, independent pulse-to-pulse errors
instead give $\tau_{\rm noise}\sim\sigma_\delta^{-2}$.  When both are
present, as illustrated for $\bar\delta=0.03$ in Fig.~\ref{fig:results}(d),
the dynamics crosses over between these two limits, so sufficiently strong
stochastic fluctuations cut off the exponentially long detuning-controlled
lifetime.  Errors in other pulse parameters and
environmental decoherence can provide additional experimental cutoffs and
are discussed in the Supplemental Material~\cite{SM}.

\emph{Experimental realization.---}
The Fibonacci implementation requires only two calibrated single-qubit
rotations and classical generation of the binary pulse sequence.  For the family
defined above, the control parameter $s$ changes both the angle and axis of
$A_s$, while $B$ remains fixed.  A resonance at $s=s_*$ can therefore be
crossed experimentally with a single control parameter,
$\delta=s-s_*$.  A natural platform is a single nitrogen-vacancy (NV) center
in diamond.  Fibonacci driving of a single NV electronic spin with
state-resolved measurements has already been demonstrated
~\cite{Liu2026}, and finite-order HSE has been measured on the same type of
platform~\cite{Pan2026}.  The required single-spin control and tomography
build on established NV-center techniques
~\cite{Dutt2007,deLange2010,Taminiau2014}.

State tomography directly provides the Bloch vectors $\bm n_t$ entering
Eq.~\eqref{eq:observable}.  To reconstruct $\bm n_t$, the spin is repeatedly
prepared in the same initial state, the first $t$ pulses of the aperiodic
sequence are applied, and the three Bloch-vector components are measured.
Repeating this procedure for successive pulse numbers gives the temporal
trajectory from which all $R_\ell(T)$ are computed.  The moment-selective
signature is then obtained from a single data set: the amplitudes
$R_\ell(T)$ with lower $\ell<\ell_*$ become small while the selected amplitude
$R_{\ell_*}(T)$ remains appreciable.  Measurements from an additional generic
initial state can be used to verify that the observed signal is not suppressed
or enhanced accidentally by the chosen preparation.

The exponentially long detuning-controlled lifetime need not be reached
directly.  Instead, one can probe the same scaling through a finite-time scan
across the resonance.  For a fixed detection threshold $\epsilon$, define
$\delta_c(T;\epsilon)$ as the magnitude of the static detuning at which
$R_{\ell_*}(T)$ falls to $\epsilon$ after an observation time $T$.  Equivalently,
$\delta_c$ gives the half-width of the finite-time region over which the
selected non-Haar signal remains detectable.  From
$\log\tau_{k_*,\epsilon}\sim\delta^{-2}$, one obtains
\begin{equation}
 |\delta_c(T;\epsilon)|\propto(\log T)^{-1/2}.
 \label{eq:width}
\end{equation}
Thus measurements at increasing pulse number should reveal a progressively
narrower long-memory region around the finite-group resonance.  The same
prediction can be tested by plotting data obtained at different $T$ against
the scaling variable $\delta^2\log T$.

Pulse-to-pulse fluctuations of the control parameter provide a distinct and
experimentally relevant cutoff.  Writing the detuning on each application of
the tunable pulse as
$\delta_t=\bar\delta+\epsilon_t$, with zero-mean independent fluctuations of
root-mean-square magnitude $\sigma_\delta$, the selected memory is limited by
$\tau_{\rm noise}\sim\sigma_\delta^{-2}$ rather than by an exponentially long
lifetime.  An experimental protocol can therefore separate the two effects:
a static scan of the mean detuning $\bar\delta$ tests
Eq.~\eqref{eq:width}, whereas deliberately increasing
$\sigma_\delta$ tests the independent-noise scaling.  Environmental
decoherence introduces an additional physical-time cutoff.  Observing the
detuning-controlled narrowing therefore requires the corresponding lifetime
to remain shorter than these noise and coherence limits \cite{SM}.

\emph{Conclusions \& outlook.---}
We have introduced a general mechanism by which hierarchically generated
aperiodic drives can separate the relaxation times of different statistical
moments of a temporal ensemble.  At a finite-group resonance, lower-order
moments can already reproduce Haar behavior while a selected higher-order
deviation remains nonzero for a parametrically longer time.  As illustrative
qubit realizations, tetrahedral and icosahedral resonances select the first
retained non-Haar moments at $k_*=3$ and $6$, respectively.  A small static
detuning from resonance produces an exponentially long lifetime whose
exponent scales inversely with the square of the detuning.  The exact Fibonacci
and silver-mean realizations further demonstrate the mechanism for distinct
aperiodic recursions.

The resulting resource is a controllable hierarchy of design lifetimes in a
temporal ensemble rather than storage of an arbitrary quantum state.  During the moment-selective
window, observables probing only moments below
$k_*$ see an approximately Haar-like temporal ensemble, while
$R_{\ell_*}(T)$ reveals the deliberately retained higher-order deviation.
This separation provides a direct way to
distinguish loss of low-order structure from loss of the selected
higher-order component, and suggests temporal state ensembles whose degree of
apparent randomness depends on the statistical order at which they are
probed~\cite{Scott2006,AmbainisEmerson2007,RoyScott2007}.

The same moment-selective dynamics can also help distinguish different control
imperfections.  A systematic calibration offset shifts the location of the
finite-group resonance, while the finite-time resonance width in
Eq.~\eqref{eq:width} reflects the sensitivity to a static detuning from that
resonance.  By contrast, independent pulse-to-pulse fluctuations introduce
the distinct noise-controlled cutoff
$\tau_{\rm noise}\sim\sigma_\delta^{-2}$.  Measurements of the resonance
position, width, and decay with pulse-to-pulse noise therefore provide
complementary information about systematic and stochastic control errors.

More broadly, moment-selective temporal designs separate two notions that are
usually discussed together in quantum-design generation: the moment order through which an
ensemble appears Haar-like and the time scale over which the first remaining
non-Haar component disappears.  Extending this separation to higher-dimensional
systems, other finite groups, and unitary rather than state moments may provide
additional routes to engineer statistical relaxation in driven quantum
systems.  Deterministic aperiodicity can therefore be used not only to generate
complex temporal ensembles, but also to control the order-by-order time scales
on which those ensembles approach Haar statistics.

\emph{Acknowledgment.}--- This work is supported by the US National Science Foundation (NSF) Grants PHY-2216774
and DMR-2406524. ChatGPT 5.6 Sol
(OpenAI) is used for assistance with computations, writing, and figure generation. 
The author takes full responsibility for the correctness and scientific value of the work. 

\bibliographystyle{apsrev4-2}
\bibliography{references}

@article{Nandy2018,
  author = {Nandy, Sourav and Sen, Arnab and Sen, Diptiman},
  title = {Steady states of a quasiperiodically driven integrable system},
  journal = {Phys. Rev. B},
  volume = {98},
  pages = {245144},
  year = {2018},
  month = {Dec},
  doi = {10.1103/PhysRevB.98.245144},
  publisher = {American Physical Society}
}

@article{Maity2019,
  author = {Maity, Somnath and Bhattacharya, Utso and Dutta, Amit and Sen, Diptiman},
  title = {Fibonacci steady states in a driven integrable quantum system},
  journal = {Phys. Rev. B},
  volume = {99},
  pages = {020306},
  year = {2019},
  month = {Jan},
  doi = {10.1103/PhysRevB.99.020306},
  publisher = {American Physical Society}
}

@article{Schmid2025,
  author = {Schmid, Harald and Peng, Yang and Refael, Gil and {von Oppen}, Felix},
  title = {Self-Similar Phase Diagram of the Fibonacci-Driven Quantum Ising Model},
  journal = {Phys. Rev. Lett.},
  volume = {134},
  pages = {240404},
  year = {2025},
  month = {Jun},
  doi = {10.1103/hn66-j8pt},
  publisher = {American Physical Society}
}

@article{Abanin2015,
  author = {Abanin, Dmitry A. and De Roeck, Wojciech and Huveneers, Francois},
  title = {Exponentially Slow Heating in Periodically Driven Many-Body Systems},
  journal = {Physical Review Letters}, volume = {115}, pages = {256803}, year = {2015},
  doi = {10.1103/PhysRevLett.115.256803}
}

@article{Mori2016,
  author = {Mori, Takashi and Kuwahara, Tomotaka and Saito, Keiji},
  title = {Rigorous Bound on Energy Absorption and Generic Relaxation in Periodically Driven Quantum Systems},
  journal = {Physical Review Letters}, volume = {116}, pages = {120401}, year = {2016},
  doi = {10.1103/PhysRevLett.116.120401}
}

@article{Abanin2017,
  author = {Abanin, Dmitry A. and De Roeck, Wojciech and Ho, Wen Wei and Huveneers, Francois},
  title = {Effective Hamiltonians, Prethermalization, and Slow Energy Absorption in Periodically Driven Many-Body Systems},
  journal = {Physical Review B}, volume = {95}, pages = {014112}, year = {2017},
  doi = {10.1103/PhysRevB.95.014112}
}

@article{Else2017,
  author = {Else, Dominic V. and Bauer, Bela and Nayak, Chetan},
  title = {Prethermal Phases of Matter Protected by Time-Translation Symmetry},
  journal = {Physical Review X}, volume = {7}, pages = {011026}, year = {2017},
  doi = {10.1103/PhysRevX.7.011026}
}

@article{DAlessio2014,
  author = {D'Alessio, Luca and Rigol, Marcos},
  title = {Long-Time Behavior of Isolated Periodically Driven Interacting Lattice Systems},
  journal = {Physical Review X}, volume = {4}, pages = {041048}, year = {2014},
  doi = {10.1103/PhysRevX.4.041048}
}

@article{Lazarides2014,
  author = {Lazarides, Achilleas and Das, Arnab and Moessner, Roderich},
  title = {Equilibrium States of Generic Quantum Systems Subject to Periodic Driving},
  journal = {Physical Review E}, volume = {90}, pages = {012110}, year = {2014},
  doi = {10.1103/PhysRevE.90.012110}
}

@article{Ponte2015,
  author = {Ponte, Pedro and Chandran, Anushya and Papic, Z. and Abanin, Dmitry A.},
  title = {Periodically Driven Ergodic and Many-Body Localized Quantum Systems},
  journal = {Annals of Physics}, volume = {353}, pages = {196--204}, year = {2015},
  doi = {10.1016/j.aop.2014.11.008}
}

@article{Dumitrescu2018,
  author = {Dumitrescu, Philipp T. and Vasseur, Romain and Potter, Andrew C.},
  title = {Logarithmically Slow Relaxation in Quasiperiodically Driven Random Spin Chains},
  journal = {Physical Review Letters}, volume = {120}, pages = {070602}, year = {2018},
  doi = {10.1103/PhysRevLett.120.070602}
}

@article{Else2020,
  author = {Else, Dominic V. and Ho, Wen Wei and Dumitrescu, Philipp T.},
  title = {Long-Lived Interacting Phases of Matter Protected by Multiple Time-Translation Symmetries in Quasiperiodically Driven Systems},
  journal = {Physical Review X}, volume = {10}, pages = {021032}, year = {2020},
  doi = {10.1103/PhysRevX.10.021032}
}

@article{Crowley2019,
  author = {Crowley, P. J. D. and Martin, I. and Chandran, A.},
  title = {Topological Classification of Quasiperiodically Driven Quantum Systems},
  journal = {Physical Review B}, volume = {99}, pages = {064306}, year = {2019},
  doi = {10.1103/PhysRevB.99.064306}
}

@article{Zhao2021,
  author = {Zhao, Hongzheng and Mintert, Florian and Moessner, Roderich and Knolle, Johannes},
  title = {Random Multipolar Driving: Tunably Slow Heating through Spectral Engineering},
  journal = {Physical Review Letters}, volume = {126}, pages = {040601}, year = {2021},
  doi = {10.1103/PhysRevLett.126.040601}
}

@article{Lapierre2020,
  author = {Lapierre, Bastien and Choo, Kenny and Tauber, Clement and Tiwari, Apoorv and Neupert, Titus and Chitra, Ramasubramanian},
  title = {Emergent Black Hole Dynamics in Critical Floquet Systems},
  journal = {Physical Review Research}, volume = {2}, pages = {023085}, year = {2020},
  doi = {10.1103/PhysRevResearch.2.023085}
}

@article{Romanelli2009,
  author = {Romanelli, A.},
  title = {The Fibonacci Quantum Walk and Its Classical Trace Map},
  journal = {Physica A}, volume = {388}, pages = {3985--3990}, year = {2009},
  doi = {10.1016/j.physa.2009.06.022}
}

@article{KolarAli1990,
  author = {Kolar, M. and Ali, M. K.},
  title = {One-Dimensional Generalized Fibonacci Tilings},
  journal = {Physical Review B}, volume = {41}, pages = {7108--7112}, year = {1990},
  doi = {10.1103/PhysRevB.41.7108}
}

@article{BaakeGrimmJoseph1993,
  author = {Baake, Michael and Grimm, Uwe and Joseph, D.},
  title = {Trace Maps, Invariants, and Some of Their Applications},
  journal = {International Journal of Modern Physics B}, volume = {7}, pages = {1527--1550}, year = {1993},
  doi = {10.1142/S021797929300247X}
}

@article{Pilatowsky2023,
  author = {Pilatowsky-Cameo, Saul and Dag, Ceren B. and Ho, Wen Wei and Choi, Soonwon},
  title = {Complete Hilbert-Space Ergodicity in Quantum Dynamics of Generalized Fibonacci Drives},
  journal = {Physical Review Letters}, volume = {131}, pages = {250401}, year = {2023},
  doi = {10.1103/PhysRevLett.131.250401}
}

@article{Pilatowsky2024,
  author = {Pilatowsky-Cameo, Saul and Marvian, Iman and Choi, Soonwon and Ho, Wen Wei},
  title = {Hilbert-Space Ergodicity in Driven Quantum Systems: Obstructions and Designs},
  journal = {Physical Review X}, volume = {14}, pages = {041059}, year = {2024},
  doi = {10.1103/PhysRevX.14.041059}
}

@article{Pilatowsky2025,
  author = {Pilatowsky-Cameo, Saul and Choi, Soonwon and Ho, Wen Wei},
  title = {Critically Slow Hilbert-Space Ergodicity in Quantum Morphic Drives},
  journal = {Physical Review Letters}, volume = {135}, pages = {140402}, year = {2025},
  doi = {10.1103/dmfd-lgcq}
}

@article{Liu2026,
  author = {Liu, Wenquan and Pan, Zou-Wei and Fu, Yue and Ho, Wen Wei and Rong, Xing},
  title = {Observation of Hierarchy of Hilbert Space Ergodicities in the Quantum Dynamics of a Single Spin},
  journal = {Physical Review Letters}, volume = {136}, pages = {020401}, year = {2026},
  doi = {10.1103/6msb-cxbc}
}

@misc{Pan2026,
  author = {Pan, Zou-Wei and Liu, Wenquan and Rong, Xing},
  title = {Experimental Investigation of Tunable-Order Hilbert-Space Ergodicity},
  year = {2026}, eprint = {2608.21959}, archivePrefix = {arXiv}, primaryClass = {quant-ph}
}

@article{Shaw2025,
  author = {Shaw, Adam L. and Mark, Daniel K. and Choi, Joonhee and Finkelstein, Ran and Scholl, Pascal and Choi, Soonwon and Endres, Manuel},
  title = {Experimental Signatures of Hilbert-Space Ergodicity: Universal Bitstring Distributions and Applications in Noise Learning},
  journal = {Physical Review X}, volume = {15}, pages = {031001}, year = {2025},
  doi = {10.1103/h6xy-zpx4}
}

@article{Mark2024,
  author = {Mark, Daniel K. and Surace, Federica and Elben, Andreas and Shaw, Adam L. and Choi, Joonhee and Refael, Gil and Endres, Manuel and Choi, Soonwon},
  title = {Maximum Entropy Principle in Deep Thermalization and in Hilbert-Space Ergodicity},
  journal = {Physical Review X}, volume = {14}, pages = {041051}, year = {2024},
  doi = {10.1103/PhysRevX.14.041051}
}

@article{Ghosh2025,
  author = {Ghosh, Souradeep and Langlett, Christopher M. and Hunter-Jones, Nicholas and Rodriguez-Nieva, Joaquin F.},
  title = {Late-Time Ensembles of Quantum States in Quantum Chaotic Systems},
  journal = {Physical Review B}, volume = {112}, pages = {094302}, year = {2025},
  doi = {10.1103/PhysRevB.112.094302}
}

@article{Logaric2025,
  author = {Logaric, Leonard and Goold, John and Dooley, Shane},
  title = {Hilbert Subspace Ergodicity},
  journal = {Physical Review B}, volume = {111}, pages = {144310}, year = {2025},
  doi = {10.1103/PhysRevB.111.144310}
}

@misc{Zhou2026,
  author = {Zhou, Yi-Neng and Zhou, Tian-Gang and Sonner, Julian},
  title = {Three Hamiltonians Are Sufficient for Unitary {$k$}-Design in Temporal Ensemble},
  year = {2026}, eprint = {2604.04205}, archivePrefix = {arXiv}, primaryClass = {quant-ph}
}

@article{Gross2007,
  author = {Gross, David and Audenaert, Koenraad and Eisert, Jens},
  title = {Evenly Distributed Unitaries: On the Structure of Unitary Designs},
  journal = {Journal of Mathematical Physics}, volume = {48}, pages = {052104}, year = {2007},
  doi = {10.1063/1.2716992}
}

@article{RoyScott2009,
  author = {Roy, Aidan and Scott, A. J.},
  title = {Unitary Designs and Codes},
  journal = {Designs, Codes and Cryptography}, volume = {53}, pages = {13--31}, year = {2009},
  doi = {10.1007/s10623-009-9290-2}
}

@article{ViolaLloyd1998,
  author = {Viola, Lorenza and Lloyd, Seth},
  title = {Dynamical Suppression of Decoherence in Two-State Quantum Systems},
  journal = {Physical Review A}, volume = {58}, pages = {2733--2744}, year = {1998},
  doi = {10.1103/PhysRevA.58.2733}
}

@article{Uhrig2007,
  author = {Uhrig, Goetz S.},
  title = {Keeping a Quantum Bit Alive by Optimized {$\pi$}-Pulse Sequences},
  journal = {Physical Review Letters}, volume = {98}, pages = {100504}, year = {2007},
  doi = {10.1103/PhysRevLett.98.100504}
}

@article{Cywinski2008,
  author = {Cywinski, Lukasz and Lutchyn, Roman M. and Nave, Cody P. and Das Sarma, S.},
  title = {How to Enhance Dephasing Time in Superconducting Qubits},
  journal = {Physical Review B}, volume = {77}, pages = {174509}, year = {2008},
  doi = {10.1103/PhysRevB.77.174509}
}

@article{Dutt2007,
  author = {Dutt, M. V. Gurudev and Childress, L. and Jiang, L. and Togan, E. and Maze, J. and Jelezko, F. and Zibrov, A. S. and Hemmer, P. R. and Lukin, M. D.},
  title = {Quantum Register Based on Individual Electronic and Nuclear Spin Qubits in Diamond},
  journal = {Science}, volume = {316}, pages = {1312--1316}, year = {2007},
  doi = {10.1126/science.1139831}
}

@article{deLange2010,
  author = {de Lange, G. and Wang, Z. H. and Riste, D. and Dobrovitski, V. V. and Hanson, R.},
  title = {Universal Dynamical Decoupling of a Single Solid-State Spin from a Spin Bath},
  journal = {Science}, volume = {330}, pages = {60--63}, year = {2010},
  doi = {10.1126/science.1192739}
}

@article{Taminiau2014,
  author = {Taminiau, T. H. and Cramer, J. and van der Sar, T. and Dobrovitski, V. V. and Hanson, R.},
  title = {Universal Control and Error Correction in Multi-Qubit Spin Registers in Diamond},
  journal = {Nature Nanotechnology}, volume = {9}, pages = {171--176}, year = {2014},
  doi = {10.1038/nnano.2014.2}
}

@article{DumontThomas1989,
  author = {Dumont, Jean-Marie and Thomas, Alain},
  title = {Systemes de Numeration et Fonctions Fractales Relatifs aux Substitutions},
  journal = {Theoretical Computer Science}, volume = {65}, pages = {153--169}, year = {1989},
  doi = {10.1016/0304-3975(89)90041-8}
}

@article{Bannai2015,
  author  = {Bannai, Eiichi and Okuda, Takayuki and Tagami, Makoto},
  title   = {Spherical Designs of Harmonic Index $t$},
  journal = {Journal of Approximation Theory},
  volume  = {195},
  pages   = {1--18},
  year    = {2015},
  doi     = {10.1016/j.jat.2014.06.010}
}

@article{Delsarte1977,
  author  = {Delsarte, P. and Goethals, J. M. and Seidel, J. J.},
  title   = {Spherical Codes and Designs},
  journal = {Geometriae Dedicata},
  volume  = {6},
  pages   = {363--388},
  year    = {1977},
  doi     = {10.1007/BF03187604}
}

@article{Reznick1995,
  author  = {Reznick, Bruce},
  title   = {Some Constructions of Spherical 5-Designs},
  journal = {Linear Algebra and its Applications},
  volume  = {226--228},
  pages   = {163--196},
  year    = {1995},
  doi     = {10.1016/0024-3795(95)00101-V}
}

@article{Mo2025,
  author = {Mo, Liang-Hong and Moessner, Roderich and Zhao, Hongzheng},
  title = {Complex and Tunable Heating in Conformal Field Theories with Structured Drives via Classical Ergodicity Breaking},
  journal = {Physical Review B}, volume = {112}, pages = {184304}, year = {2025},
  doi = {10.1103/dvlw-hl7t}
}

@article{LiuNature2026,
  author = {Liu, Zheng-He and others},
  title = {Prethermalization by Random Multipolar Driving on a 78-Qubit Processor},
  journal = {Nature}, volume = {650}, pages = {79--85}, year = {2026},
  doi = {10.1038/s41586-025-09977-x}
}

@article{GrimusLudl2012,
  author = {Grimus, Walter and Ludl, Patrick O.}, title = {Finite Flavour Groups of Fermions},
  journal = {Journal of Physics A: Mathematical and Theoretical}, volume = {45}, pages = {233001}, year = {2012}, doi = {10.1088/1751-8113/45/23/233001}
}

@misc{SM,
  title = {{See Supplemental Material appended below for the general moment formulation, finite-dimensional recursion for recursively generated aperiodic drives, detuning scaling, exact Fibonacci and silver-mean constructions, and additional experimental imperfections.}}
}

@article{Scott2006,
  author = {Scott, A. J.},
  title = {Tight Informationally Complete Quantum Measurements},
  journal = {Journal of Physics A: Mathematical and General},
  volume = {39}, pages = {13507--13530}, year = {2006},
  doi = {10.1088/0305-4470/39/43/009}
}

@inproceedings{AmbainisEmerson2007,
  author = {Ambainis, Andris and Emerson, Joseph},
  title = {Quantum {$t$}-Designs: {$t$}-Wise Independence in the Quantum World},
  booktitle = {Twenty-Second Annual IEEE Conference on Computational Complexity (CCC'07)},
  pages = {129--140}, year = {2007},
  doi = {10.1109/CCC.2007.26}
}

@article{RoyScott2007,
  author = {Roy, Aidan and Scott, A. J.},
  title = {Weighted Complex Projective 2-Designs from Bases: Optimal State Determination by Orthogonal Measurements},
  journal = {Journal of Mathematical Physics}, volume = {48}, pages = {072110}, year = {2007},
  doi = {10.1063/1.2748617}
}

\clearpage
\begin{widetext}
\allowdisplaybreaks
\setcounter{section}{0}
\setcounter{subsection}{0}
\setcounter{equation}{0}
\setcounter{theorem}{0}
\setcounter{secnumdepth}{2}
\renewcommand{\thesection}{S\arabic{section}}
\renewcommand{\thesubsection}{\thesection.\arabic{subsection}}
\hypersetup{hidelinks}

\begin{center}
{\large\bfseries Supplemental Material for ``Moment-Selective Quantum Designs in Aperiodic Temporal Ensembles''\par}
\vspace{0.8em}
{\normalsize Yang Peng\par}
\vspace{0.35em}
{\small Department of Physics and Astronomy, California State University, Northridge, Northridge, California 91330, USA\par}
{\small Institute of Quantum Information and Matter and Department of Physics, California Institute of Technology, Pasadena, CA 91125, USA\par}
\end{center}
\vspace{1.0em}

This Supplemental Material provides the derivations deferred from the main
text.  It is organized around five questions: (i) how the state-moment
formulation extends beyond qubits; (ii) how recursively generated aperiodic
drives produce a finite-dimensional recursion for temporal averages; (iii) why
a smooth finite-period return of that recursion gives a quadratic, or more
generally even-power, detuning dependence; (iv) how the Fibonacci and
silver-mean realizations produce the finite-group resonances and the period-12
and period-20 returns used in the analysis; and (v) how additional control
errors and environmental decoherence limit the experimentally observable
window.  

\section{Moment-selective temporal designs beyond a qubit}\label{sec:general-moments}

We first state the moment language in a form that applies to an arbitrary
 finite-dimensional quantum system.  Let $\mathcal H\cong\mathbb C^d$ be the
 Hilbert space and let
\begin{equation}
 |\psi_t\rangle=U(t)|\psi_0\rangle
\end{equation}
be the pure state reached after $t$ pulses.  The order-$k$ temporal state moment
 is
\begin{equation}
 M_k(T)=\frac1T\sum_{t=0}^{T-1}
 (|\psi_t\rangle\langle\psi_t|)^{\otimes k}.
 \label{eq:sm-state-moment}
\end{equation}
The corresponding Haar moment, denoted $M_k^{\rm Haar}$, is obtained by
 averaging the same operator over the unitarily invariant distribution of pure
 states.  A temporal ensemble is a state $k$-design when
$M_j(T)=M_j^{\rm Haar}$ for every $j\le k$.  More generally, we call the
temporal ensemble a \emph{moment-selective temporal design} over a given
time window when lower moments are already Haar-like while a selected higher
moment remains detectably non-Haar.

For the representation-theoretic form of this statement, define
 $W_k=\Sym^k(\mathcal H)$, the symmetric $k$-copy Hilbert space, and let
 $D_k(U)=\Sym^k(U)$ be the action of a unitary $U$ on $W_k$. 
We define the time-averaged conjugation map 
\begin{equation}
 \mathcal T_T^{(k)}(X)=\frac1T\sum_{t=0}^{T-1}
 D_k[U(t)]X D_k[U(t)]^\dagger,
 \qquad X\in\operatorname{End}(W_k).
 \label{eq:sm-state-twirl}
\end{equation}
The group $G$ acts on operators $X\in\operatorname{End}(W_k)$ by conjugation,
\begin{equation}
 X\mapsto D_k(g) X D_k(g)^\dagger,
 \qquad g\in G,
\end{equation}
where $\operatorname{End}(W_k)$ denotes the vector space of linear operators on
$W_k$.  An operator is called $G$-invariant if it is unchanged by this action
for every $g\in G$.

Because $\operatorname{End}(W_k)$ is finite dimensional, it can be decomposed
into the invariant sector and a finite collection of sectors that transform
nontrivially under $G$:
\begin{equation}
 \operatorname{End}(W_k)
 \cong
 \operatorname{Fix}_G\!\left[\operatorname{End}(W_k)\right]
 \oplus
 \bigoplus_{\rho\neq 1}
 \mathbb C^{m_{k,\rho}}\otimes V_\rho .
 \label{eq:sm-irrep-decomp}
\end{equation}
Here
$\operatorname{Fix}_G[\operatorname{End}(W_k)]$ is the subspace of operators
that are invariant under every element of $G$.  Each $V_\rho$ is a subspace
that transforms according to a nontrivial irreducible representation $\rho$:
within this sector, applying $g\in G$ acts through a unitary matrix $\rho(g)$.
The integer $m_{k,\rho}$ counts how many copies of the same transformation
sector occur.  The symbol $\rho=1$ denotes the trivial representation, for
which $\rho(g)=1$ for every $g\in G$.

Haar averaging over $G$ leaves the invariant sector unchanged and averages
every nontrivial sector to zero.  Therefore, the temporal order-$k$ average
generated by the sequence $U(t)$ approaches the corresponding Haar average
precisely when
\begin{equation}
 \frac{1}{T}\sum_{t=0}^{T-1}\rho[U(t)]
 \longrightarrow 0
 \qquad (T\to\infty)
 \label{eq:sm-fourier-decay}
\end{equation}
for every nontrivial representation sector $\rho$ that occurs in
Eq.~\eqref{eq:sm-irrep-decomp}.  This is the group-theoretic analogue of
Fourier averaging on a circle: a uniform average keeps the constant mode and
removes all nonzero Fourier modes.

A finite subgroup $H\subset G$ can reproduce the Haar state moments up to a
given order even though $H$ contains only finitely many elements.  To state the
condition precisely, consider one of the nontrivial representation sectors
$V_\rho$ introduced above.  Its $H$-invariant subspace is
\begin{equation}
 \operatorname{Fix}_{\rho}(H)
 =
 \{v\in V_\rho:\rho(h)v=v
 \text{ for every }h\in H\}.
 \label{eq:sm-fixed-space}
\end{equation}
Thus, $\operatorname{Fix}_{\rho}(H)$ contains the components in this sector
that remain unchanged under every element of $H$.

Suppose that, at long times, the temporal products $U(t)$ sample the elements
of $H$ uniformly.  Averaging over the trajectory is then equivalent to
averaging uniformly over $H$.  Such an average agrees with the $G$-Haar
average through state-moment order $k$ if and only if no nontrivial
representation sector appearing through that order contains an
$H$-invariant component:
\begin{equation}
 \operatorname{Fix}_{\rho}(H)=\{0\}
 \qquad
 \text{for every nontrivial $\rho$ appearing through order $k$}.
 \label{eq:sm-subgroup-design}
\end{equation}
In other words, the subgroup average removes all non-Haar components that can
contribute up to order $k$.  A finite set of states with this property is
called a projective state $k$-design~\cite{Gross2007,RoyScott2009}.

For a qubit, the relevant continuous group is $G=SU(2)$.  The symmetric
$k$-copy space $W_k=\operatorname{Sym}^k(\mathbb C^2)$ transforms as a
spin-$k/2$ system.  The corresponding action on operators decomposes as
\begin{equation}
 D^{(k/2)}\otimes D^{(k/2)*}
 \cong
 \bigoplus_{\ell=0}^{k}D^{(\ell)},
 \label{eq:sm-qubit-decomp}
\end{equation}
where $D^{(\ell)}$ denotes the spin-$\ell$ representation of $SU(2)$.  On the
Bloch sphere, the same sectors are represented by spherical harmonics of
degree $\ell$.  The $\ell=0$ sector is rotationally invariant and corresponds
to the Haar contribution, whereas each $\ell>0$ sector measures a possible
angular anisotropy of the state distribution.

The quantity $R_\ell(T)$ used in the main text is the rotationally invariant norm
of the degree-$\ell$ component of the temporal state distribution.  Hence
$R_\ell(T)=0$ means that no degree-$\ell$ anisotropy remains.  If a finite
subgroup $H$ has no invariant component for
$\ell=1,\ldots,\ell_*-1$, but does have one at $\ell=\ell_*$, then the subgroup
average agrees with the Haar state moments through order
$k=\ell_*-1$ and can first differ from Haar at order
\begin{equation}
 k_*=\ell_*.
\end{equation}
This is the origin of the moment-selective temporal-design behavior shown in
Fig.~1(c) of the main text: for the binary tetrahedral group $2T$, the first
retained harmonic is
$\ell_*=3$, whereas for the binary icosahedral group $2I$, it is
$\ell_*=6$.

\section{Finite-dimensional recursion for recursively generated drives}
\label{sec:transfer}

The main text considers recursively generated aperiodic pulse sequences, with the
Fibonacci and silver-mean drives as the principal examples.  We first introduce
the notation needed to describe these sequences.

An \emph{alphabet} $\mathcal A$ is a finite set of symbols.  For example, for
a drive constructed from two elementary unitary gates one may take
$\mathcal A=\{\mathsf A,\mathsf B\}$, where $\mathsf A$ and $\mathsf B$ label the two gates (unitary pulses).  A finite word
over $\mathcal A$ is a finite ordered string of these symbols, such as
$\mathsf A$, $\mathsf B\mathsf A$, or $\mathsf B\mathsf A\mathsf B\mathsf B\mathsf A$.

A \emph{substitution} $\sigma$ specifies how each letter is replaced by a
nonempty finite word.  We choose the Fibonacci convention
\begin{equation}
 \sigma(\mathsf A)=\mathsf B,
 \qquad
 \sigma(\mathsf B)=\mathsf B\mathsf A.
 \label{eq:sm-fib-substitution}
\end{equation}
Repeated substitution starting from $\mathsf B$ gives
\[
 \mathsf B
 \;\longrightarrow\;
 \mathsf B\mathsf A
 \;\longrightarrow\;
 \mathsf B\mathsf A\mathsf B
 \;\longrightarrow\;
 \mathsf B\mathsf A\mathsf B\mathsf B\mathsf A
 \;\longrightarrow\;\cdots .
\]
This is the symbolic version of the Fibonacci pulse ordering used in the
numerical simulations and in the unitary-block recursion of the main text.

The only general property of the substitution that we need is that repeated
substitution mixes all letters and produces pulse blocks with a common
exponential growth rate.  In standard terminology, such a substitution is
\emph{primitive}: there exists an integer $m$ such that every letter in
$\mathcal A$ occurs in $\sigma^m(a)$ for every starting letter
$a\in\mathcal A$.  For a primitive growing substitution, the length of the
$n$th pulse block grows asymptotically by a common factor $\lambda>1$,
\begin{equation}
 \frac{|\sigma^{n+1}(a)|}{|\sigma^n(a)|}
 \longrightarrow
 \lambda
 \qquad (n\to\infty)
 \label{eq:sm-inflation}
\end{equation}
for every starting letter $a$.  Here $|\sigma^n(a)|$ is the number of
elementary pulses in the $n$th pulse block generated from $a$.  For the
Fibonacci substitution, $\lambda=\varphi=(1+\sqrt5)/2$.

We next describe the unitary evolution generated by a word.  Associate each
letter $a\in\mathcal A$ with a unitary gate $U_a$.  For a finite word
$x=x_1x_2\cdots x_{|x|}$, let $P(x)$ denote the physical unitary obtained by
applying the corresponding gates in chronological order.  We use the
convention that later gates multiply on the left.  Therefore, if the pulse
block $x$ is followed by the pulse block $y$,
\begin{equation}
 P(xy)=P(y)P(x).
 \label{eq:sm-product-convention}
\end{equation}
For example, $P(\mathsf A\mathsf B)=U_{\mathsf B}U_{\mathsf A}$.

Let $G$ be the group generated by the elementary gates, and let $\rho$ be a
finite-dimensional unitary representation of $G$.  For a word $x$, define
the representation-valued temporal average
\begin{equation}
 \overline S_\rho(x)
 =
 \frac{1}{|x|}
 \sum_{t=0}^{|x|-1}
 \rho\!\left[
 P\!\left(x_{[0,t)}\right)
 \right],
 \label{eq:sm-prefix-average}
\end{equation}
where $x_{[0,t)}$ denotes the first $t$ letters of $x$.  The case $t=0$
corresponds to the empty prefix, for which $P(x_{[0,0)})=I$.
Equation~\eqref{eq:sm-prefix-average} is the finite-pulse-block version of
the temporal representation average
$T^{-1}\sum_{t=0}^{T-1}\rho[U(t)]$ introduced above.

The temporal average has a simple exact rule under concatenation.  If $x$ is
followed by $y$, then
\begin{equation}
 \overline S_\rho(xy)
 =
 \frac{|x|}{|x|+|y|}\,
 \overline S_\rho(x)
 +
 \frac{|y|}{|x|+|y|}\,
 \overline S_\rho(y)\rho[P(x)].
 \label{eq:sm-prefix-concat}
\end{equation}
The first term contains the prefixes that lie inside $x$.  For a prefix that
extends into $y$, the complete block $x$ has already been applied, which
produces the factor $\rho[P(x)]$ multiplying the second term.  The two
coefficients are simply the fractions of pulses belonging to the two
constituent blocks and sum to one.

For the Fibonacci drive, Eq.~\eqref{eq:sm-fib-substitution} gives
\begin{equation}
 \sigma^{n+1}(\mathsf A)
 =
 \sigma^n(\mathsf B),
 \qquad
 \sigma^{n+1}(\mathsf B)
 =
 \sigma^n(\mathsf B)\sigma^n(\mathsf A).
 \label{eq:sm-fib-block-recursion}
\end{equation}
Thus the temporal averages of successive pulse blocks obey
\begin{align}
 \overline S_\rho[\sigma^{n+1}(\mathsf B)]
 &=
 \frac{|\sigma^n(\mathsf B)|}{|\sigma^{n+1}(\mathsf B)|}
 \overline S_\rho[\sigma^n(\mathsf B)]
 \nonumber\\
 &\quad+
 \frac{|\sigma^n(\mathsf A)|}{|\sigma^{n+1}(\mathsf B)|}
 \overline S_\rho[\sigma^n(\mathsf A)]
 \rho[P(\sigma^n(\mathsf B))],
 \label{eq:sm-fib-prefix-B}\\
 \overline S_\rho[\sigma^{n+1}(\mathsf A)]
 &=
 \overline S_\rho[\sigma^n(\mathsf B)].
 \label{eq:sm-fib-prefix-A}
\end{align}
The $(n+1)$th pulse block generated from $\mathsf B$ consists of an $n$th
block generated from $\mathsf B$ followed by one generated from $\mathsf A$.
The second contribution is therefore multiplied by
$\rho[P(\sigma^n(\mathsf B))]$, the unitary accumulated during the first
constituent block.

The same construction applies to a general substitution.  If
\[
 \sigma(a)=b_{a,1}b_{a,2}\cdots b_{a,k_a},
\]
then
\begin{equation}
 \sigma^{n+1}(a)
 =
 \sigma^n(b_{a,1})
 \sigma^n(b_{a,2})
 \cdots
 \sigma^n(b_{a,k_a}).
 \label{eq:sm-general-block-recursion}
\end{equation}
Repeated use of Eq.~\eqref{eq:sm-prefix-concat} gives
\begin{align}
 \overline S_\rho[\sigma^{n+1}(a)]
 &=
 \sum_{j=1}^{k_a}
 \frac{|\sigma^n(b_{a,j})|}
      {|\sigma^{n+1}(a)|}\,
 \overline S_\rho[\sigma^n(b_{a,j})]
 \nonumber\\
 &\quad\times
 \rho\!\left[
 P\!\left(
 \sigma^n(b_{a,1})\cdots
 \sigma^n(b_{a,j-1})
 \right)
 \right].
 \label{eq:sm-general-prefix-recursion}
\end{align}
For $j=1$, the product inside $P$ is empty and is understood to be the
identity.  The unitary in the second line is simply the evolution accumulated
before the $j$th constituent pulse block begins.

The coefficients in Eq.~\eqref{eq:sm-general-prefix-recursion} are positive
and sum exactly to one because pulse-block lengths add under concatenation:
\begin{equation}
 |\sigma^{n+1}(a)|
 =
 \sum_{j=1}^{k_a}|\sigma^n(b_{a,j})|.
 \label{eq:sm-length-addition}
\end{equation}
Thus the recursion expresses the temporal average of an $(n+1)$th pulse block
as a weighted average of the temporal averages of its constituent $n$th pulse
blocks, after accounting for the unitary evolution accumulated before each
constituent block.

To quantify cancellation in this recursion, we use the Hilbert--Schmidt norm
of a matrix $A$,
\begin{equation}
 \|A\|_{\rm HS}^2
 =
 \operatorname{Tr}(A^\dagger A),
 \label{eq:sm-hs-norm}
\end{equation}
where $\operatorname{Tr}$ denotes the matrix trace.  Because $\rho$ is a
unitary representation, right multiplication by any $\rho(g)$ leaves this
norm unchanged:
\begin{equation}
 \|A\rho(g)\|_{\rm HS}=\|A\|_{\rm HS}.
 \label{eq:sm-unitary-hs}
\end{equation}

Equation~\eqref{eq:sm-general-prefix-recursion} and the triangle inequality
therefore imply
\begin{align}
 \left\|
 \overline S_\rho[\sigma^{n+1}(a)]
 \right\|_{\rm HS}
 &\le
 \sum_{j=1}^{k_a}
 \frac{|\sigma^n(b_{a,j})|}
      {|\sigma^{n+1}(a)|}
 \left\|
 \overline S_\rho[\sigma^n(b_{a,j})]
 \right\|_{\rm HS}
 \nonumber\\
 &\le
 \max_{b\in\mathcal A}
 \left\|
 \overline S_\rho[\sigma^n(b)]
 \right\|_{\rm HS}.
 \label{eq:sm-single-block-contraction}
\end{align}
Taking the maximum over the starting letter $a$ gives
\begin{equation}
 \max_{a\in\mathcal A}
 \left\|
 \overline S_\rho[\sigma^{n+1}(a)]
 \right\|_{\rm HS}
 \le
 \max_{a\in\mathcal A}
 \left\|
 \overline S_\rho[\sigma^n(a)]
 \right\|_{\rm HS}.
 \label{eq:sm-contraction}
\end{equation}
Hence the recursive construction cannot increase the largest
representation-valued temporal average among the pulse blocks.  Reduction
occurs when the constituent contributions point in different directions in
the representation space and therefore partially cancel.  If all transformed
contributions entering a given pulse block are identical, no cancellation
occurs in that update.

This observation also clarifies the special role of the finite-group
resonances discussed in the main text.  At an exact resonance, suppose the
elementary gates generate a finite subgroup $H\subset G$.  Every temporal
product then belongs to $H$.  If the representation $\rho$ contains a nonzero
$H$-invariant vector $v$, defined by
\begin{equation}
 \rho(h)v=v
 \qquad
 \text{for every }h\in H,
 \label{eq:sm-H-invariant-vector}
\end{equation}
then every term in the temporal average acts identically on $v$.  Consequently,
\begin{equation}
 \overline S_\rho(x)v=v
 \label{eq:sm-retained-component}
\end{equation}
for every pulse block $x$ whose unitary products lie in $H$.  The corresponding
representation component therefore cannot decay to zero.  This is the
retained non-Haar component at the finite-group resonance.

By contrast, the absence of an $H$-invariant vector does not by itself imply
that the corresponding temporal average must decrease to zero.  Whether it
does so also depends on the order in which the group elements are visited by
the recursively generated sequence.  For the Fibonacci and silver-mean drives
considered here, we show below that, in the unwanted lower-order
representation sectors,
\begin{equation}
 \left\|
 \overline S_\rho[\sigma^n(a)]
 \right\|_{\rm HS}
 \longrightarrow 0
 \qquad (n\to\infty),
\end{equation}
whereas the first sector containing the relevant subgroup invariant retains a
nonzero component at the exact resonance.  Thus the recursively generated
trajectory reproduces the Haar moments in the lower-order sectors while
preserving the first symmetry-allowed non-Haar component.

\section{Static detuning and the even-power law}
\label{sec:leakage}

We now explain the quadratic dependence on static detuning discussed in the
main text.  Let $\delta$ denote a time-independent displacement of a control
parameter from its resonant value, with $\delta=0$ at the exact finite-group
resonance.

At resonance, the subgroup-invariant component identified in the previous
section gives a nonzero temporal average.  A static detuning changes the
unitary gates and therefore makes the contributions entering the recursive
average slightly different from one another.  Repeated averaging then reduces
the retained component as the recursion proceeds.  When this decrease is
exponential in the recursion order $n$, we define its decay rate
$\Delta_\rho$ by the following asymptotic relation
\begin{equation}
 \max_{a\in\mathcal A}
 \left\|
 \overline S_\rho[\sigma^n(a)]
 \right\|_{\rm HS}
 \sim
 e^{-n\Delta_\rho}.
 \label{eq:sm-gap}
\end{equation}
Thus $\Delta_\rho$ measures the decay per recursion step in the
representation sector $\rho$.  At the exact resonance, the retained component
does not decay and therefore $\Delta_\rho(0)=0$.  For a qubit,
$\rho=D^{(\ell)}$, and the corresponding quantity is denoted
$\Delta_\ell$ in the main text.

The origin of the quadratic dependence of $\Delta_\rho(\delta)$ on small $\delta$ follows directly from a
general property of weighted averages.  For any matrices $Y_j$ and positive
weights $w_j$ satisfying $\sum_jw_j=1$,
\begin{equation}
 \sum_j w_j\|Y_j\|_{\rm HS}^2
 -
 \left\|\sum_j w_jY_j\right\|_{\rm HS}^2
 =
 \frac{1}{2}
 \sum_{j,k}w_jw_k
 \|Y_j-Y_k\|_{\rm HS}^2.
 \label{eq:sm-variance-identity}
\end{equation}
The left-hand side of Eq.~\eqref{eq:sm-variance-identity} is the decrease in squared norm produced by averaging.
If all $Y_j$ are identical, their average is identical to each of them and
there is no decrease.  If they differ, they partially cancel.  The
right-hand side shows that the amount of this decrease is determined by the
squared differences $\|Y_j-Y_k\|_{\rm HS}^2$.

To apply Eq.~\eqref{eq:sm-variance-identity} to the recursive average, compare
it directly with Eq.~\eqref{eq:sm-general-prefix-recursion}.  For a fixed
starting letter $a$ and recursion step $n$, the weights and matrices in
Eq.~\eqref{eq:sm-variance-identity} are
\begin{equation}
 w_j=
 \frac{|\sigma^n(b_{a,j})|}
      {|\sigma^{n+1}(a)|},
 \qquad
 Y_j(\delta)=
 \overline S_\rho[\sigma^n(b_{a,j})]\,
 \rho\!\left[
 P\!\left(
 \sigma^n(b_{a,1})\cdots\sigma^n(b_{a,j-1})
 \right)
 \right],
 \label{eq:sm-Yj-definition}
\end{equation}
where all unitary gates entering $\overline S_\rho$ and $P$ are evaluated at
the detuning $\delta$.  Thus $Y_j(\delta)$ is precisely the contribution of
the $j$th constituent $n$th pulse block to the temporal average of the
$(n+1)$th pulse block generated from $a$, including the unitary evolution
accumulated before that constituent block begins.

At the exact resonance, let $v$ be the $H$-invariant vector defined in
Eq.~\eqref{eq:sm-H-invariant-vector}.  Every unitary product appearing in
Eq.~\eqref{eq:sm-Yj-definition} then belongs to $H$.  Using
Eq.~\eqref{eq:sm-retained-component}, one therefore has
\begin{equation}
 Y_j(0)v=v
 \qquad
 \text{for every }j.
 \label{eq:sm-Yj-resonance}
\end{equation}
Thus all constituent pulse blocks give the same contribution to the retained
component at resonance, and averaging them does not reduce that component.
A small static detuning makes these contributions different.  If, for at
least one pair $(j,k)$,
\begin{equation}
 [Y_j(\delta)-Y_k(\delta)]v
 =
 O(\delta)
 \label{eq:sm-linear-misalignment}
\end{equation}
with a nonzero first-order term, then the squared difference is of order
$\delta^2$.  Applying the variance identity
Eq.~\eqref{eq:sm-variance-identity} to the vectors $Y_j(\delta)v$ therefore
shows that the reduction of the retained component produced by this
recursive averaging step begins at order $\delta^2$.  Because the
right-hand side of Eq.~\eqref{eq:sm-variance-identity} is a sum of
nonnegative squared norms, the quadratic coefficient is positive whenever
at least one such first-order difference is nonzero.

To turn this reduction in one recursive update into the long-$n$ decay
rate $\Delta_\rho$, we use a finite-period return of the unitary part of the
finite-dimensional recursion.  Suppose that, along the detuning family under
consideration, the pulse-block unitaries return after a fixed number $r$ of
recursion steps, up to a simultaneous unitary conjugation.  For a primitive
substitution, the length fractions in
Eq.~\eqref{eq:sm-general-prefix-recursion} approach fixed positive values as
$n$ increases.  Therefore, in a basis that follows the common conjugation,
the $r$-step averaging map approaches a fixed finite map; the remaining
changes in the length fractions vanish asymptotically and do not change the
leading decay exponent.

We assume that the retained component is isolated, meaning that no other
independent component under consideration has zero decay rate at
$\delta=0$, and that the elementary pulse unitaries depend smoothly on
$\delta$.  The limiting $r$-step map then varies smoothly with $\delta$,
and the reduction accumulated over one return cycle has the same leading
power of $\delta$ as the squared differences in
Eq.~\eqref{eq:sm-variance-identity}.  Repeating this finite map produces the
exponential decay in Eq.~\eqref{eq:sm-gap}.  The two realizations below
satisfy precisely this condition: the physical Fibonacci family has a
period-12 trace return, Eq.~\eqref{eq:sm-fib-period12}, while the
silver-mean detuning is chosen on a period-20 trace family,
Eq.~\eqref{eq:sm-silver-branch}.

More generally, suppose that symmetry or fine tuning makes all differences
between the relevant contributions vanish through order $\delta^{q-1}$, so
that their first nonzero differences occur at order
\begin{equation}
 Y_j(\delta)-Y_k(\delta)=O(\delta^q),
 \qquad q\ge1,
 \label{eq:sm-order-q-misalignment}
\end{equation}
with at least one difference having a nonzero term of order $\delta^q$.
Equation~\eqref{eq:sm-variance-identity} then shows that the reduction in
squared norm begins at order $\delta^{2q}$.  Consequently, the decay rate has
the form
\begin{equation}
 \Delta_\rho(\delta)
 =
 c_\rho\,\delta^{2q}
 +
 O(\delta^{2q+1}),
 \qquad
 c_\rho>0.
 \label{eq:sm-even-power}
\end{equation}

For a one-parameter detuning that changes at least one of the relevant
contributions at first order, $q=1$, and therefore
\begin{equation}
 \Delta_\rho(\delta)
 =
 c_\rho\,\delta^2+O(\delta^3).
 \label{eq:sm-quadratic-gap}
\end{equation}
This is the quadratic law shown in Fig.~2(a) of the main text.  A higher
leading even power can occur only when symmetry or fine tuning removes the
lower-order differences.  The physical origin of the even power is therefore
simple: detuning produces a difference between contributions at order
$\delta^q$, while the reduction caused by averaging is proportional to the
square of that difference.

\section{From recursion order to the number of physical pulses}
\label{sec:physical-time}

Equation~\eqref{eq:sm-gap} describes how the temporal representation average
decreases with the pulse-block order $n$.  An experiment, however, measures
time by the number of physical pulses $T$.  We now relate these two
descriptions.

Equation~\eqref{eq:sm-inflation} states that the length of the $n$th pulse
block generated by a primitive growing substitution increases exponentially
with $n$.  Equivalently,
\begin{equation}
 \log |\sigma^n(a)|
 =
 n\log\lambda+O(1)
 \qquad (n\to\infty).
 \label{eq:sm-log-block-growth}
\end{equation}
Thus, when the physical pulse number $T$ is of the order of
$|\sigma^n(a)|$,
\begin{equation}
 n
 =
 \frac{\log T}{\log\lambda}+O(1).
 \label{eq:sm-n-to-T}
\end{equation}

According to Eq.~\eqref{eq:sm-gap}, the temporal representation average over
the $n$th pulse blocks decreases as $e^{-n\Delta_\rho}$.  Substituting
Eq.~\eqref{eq:sm-n-to-T} therefore gives
\begin{equation}
 e^{-n\Delta_\rho}
 \sim
 T^{-\Delta_\rho/\log\lambda},
 \label{eq:sm-block-power-law}
\end{equation}
up to a multiplicative factor that does not grow with $T$.  Thus an
exponential decrease with the pulse-block order becomes a power-law decrease
with the physical pulse number.

Equation~\eqref{eq:sm-block-power-law} was derived for pulse numbers
corresponding to the ends of recursively generated blocks of the form
$\sigma^n(a)$.  We now explain why the same asymptotic behavior also applies
at an arbitrary pulse number $T$ in the actual nonrepeated sequence.

Consider the prefix of length $T$ of the infinite pulse sequence, where a
\emph{prefix} simply means the first $T$ pulses of the sequence.  For the
Fibonacci and silver-mean recursions considered here, such a prefix can be
written as a concatenation of recursively generated pulse blocks
$\sigma^m(a)$ with successively smaller values of $m$.  For example, one
first includes the longest recursively generated pulse block that fits within
the first $T$ pulses, and then decomposes the remaining part in the same way
using shorter pulse blocks.

Applying the concatenation rule
Eq.~\eqref{eq:sm-prefix-concat} repeatedly expresses the temporal average over
the first $T$ pulses in terms of the temporal averages
$\overline S_\rho[\sigma^m(a)]$ of these recursively generated pulse blocks.
The contribution from each block is weighted by its fraction of the total
length $T$ and is multiplied on the right by a unitary representation matrix
that accounts for the evolution accumulated before that block begins.
According to Eq.~\eqref{eq:sm-unitary-hs}, this unitary factor does not change
the Hilbert--Schmidt norm.

Because $|\sigma^m(a)|$ grows exponentially with $m$, the pulse blocks with
successively smaller values of $m$ are exponentially shorter.
Therefore the asymptotic power-law dependence obtained in
Eq.~\eqref{eq:sm-block-power-law} for the recursively generated pulse blocks
also governs the temporal average at arbitrary pulse numbers in the full
nonrepeated sequence.

For a fixed small threshold $\epsilon$, we define the lifetime of the
representation sector $\rho$ as the pulse number after which its temporal
average remains below $\epsilon$:
\begin{equation}
 \tau_{\rho,\epsilon}
 =
 \min\left\{
 T_0:
 \left\|
 \frac{1}{T}\sum_{t=0}^{T-1}\rho[U(t)]
 \right\|_{\rm HS}
 \le\epsilon
 \quad\text{for every }T\ge T_0
 \right\}.
 \label{eq:sm-sector-lifetime}
\end{equation}
We assume that the slowly decaying component characterized by
$\Delta_\rho$ is present with nonzero weight in this temporal average;
otherwise that component would not determine the measured lifetime.

Using Eq.~\eqref{eq:sm-block-power-law}, reaching a fixed threshold requires
\begin{equation}
 T^{-\Delta_\rho/\log\lambda}
 \sim \epsilon.
\end{equation}
Taking the logarithm shows that, as $\Delta_\rho\to0$,
\begin{equation}
 \log\tau_{\rho,\epsilon}
 \propto
 \frac{1}{\Delta_\rho},
 \label{eq:sm-gap-to-time}
\end{equation}
where the proportionality factor depends on the fixed threshold and on the
inflation factor, but not on the leading power of the detuning.

Finally, if $\Delta_\rho$ is governed by Eq.~\eqref{eq:sm-even-power} which gives
$\Delta_\rho(\delta)\propto|\delta|^{2q}$ near resonance, we have
\begin{equation}
 \log\tau_{\rho,\epsilon}(\delta)
 \propto
 |\delta|^{-2q}.
 \label{eq:sm-essential-law}
\end{equation}
For the usual case in which the relevant differences between contributions
appear already at first order in the static detuning, $q=1$, and hence
\begin{equation}
 \log\tau_{\rho,\epsilon}(\delta)
 \propto
 |\delta|^{-2}.
 \label{eq:sm-quadratic-lifetime}
\end{equation}
Equivalently, the lifetime grows exponentially as
$\exp[\text{const.}/\delta^2]$ when the resonance is approached.

For the Fibonacci drive, $\lambda=\varphi=(1+\sqrt5)/2$, while for the
silver-mean drive considered below, $\lambda=1+\sqrt2$.  These different
inflation factors change the numerical relation between recursion order and
physical pulse number, but not the inverse-power dependence of
$\log\tau_{\rho,\epsilon}$ on the detuning.

\section{Exact Fibonacci realization}
\label{sec:fibonacci}

We now specialize the general results above to the Fibonacci drive considered
in the main text.  The symbolic Fibonacci words are defined by
Eq.~\eqref{eq:sm-fib-substitution}; here we connect those words to the physical
pulse unitaries and to the unitary-block recursion used in the main text.

For the one-parameter Fibonacci family, let $P_s(x)$ denote the unitary
produced by the pulse word $x$.  The two elementary letters are assigned
\begin{equation}
 P_s(\mathsf A)=A_s,
 \qquad
 P_s(\mathsf B)=B,
 \label{eq:sm-fib-letter-unitary}
\end{equation}
where
\begin{align}
 B
 &=
 \frac12\Id
 -
 i\frac{\sqrt3}{2}\sigma_z,
 \label{eq:sm-B}
 \\
 A_s
 &=
 s\Id
 -
 i\left[
 \sqrt{\frac{2(1+s)(1-2s)}{3}}\,\sigma_x
 -
 \frac{1+s}{\sqrt3}\sigma_z
 \right],
 \qquad
 -1<s<\frac12.
 \label{eq:sm-As}
\end{align}
Thus $\mathsf A$ and $\mathsf B$ are symbolic letters, while $A_s$ and $B$
are the corresponding $2\times2$ unitary pulse matrices.

With the chronological product convention
Eq.~\eqref{eq:sm-product-convention}, Eq.~\eqref{eq:sm-fib-block-recursion}
gives
\begin{align}
 P_s[\sigma^{n+1}(\mathsf A)]
 &=
 P_s[\sigma^n(\mathsf B)],
 \nonumber\\
 P_s[\sigma^{n+1}(\mathsf B)]
 &=
 P_s[\sigma^n(\mathsf A)]
 P_s[\sigma^n(\mathsf B)].
 \label{eq:sm-fib-product-recursion}
\end{align}
At $n=0$ the pair is $(A_s,B)$, so this is exactly the unitary-block recursion
used in the main text: if $W_0=A_s$ and $W_1=B$, then
$W_{n+2}=W_nW_{n+1}$.  The order of the matrix product follows from the fact
that later physical pulses multiply on the left.

\subsection{Period-12 return of the Fibonacci trace recursion}
\label{sec:fib-period12}

We next show that the Fibonacci control family has a finite-period recursion
in terms of three traces.  This property will also be used in the control-error
analysis below.

For an $SU(2)$ rotation $U$ through an angle $\theta$,
\begin{equation}
 \frac12\operatorname{Tr}U
 =
 \cos\frac{\theta}{2}.
 \label{eq:sm-half-trace-angle}
\end{equation}
The half trace therefore determines the rotation angle independently of the
rotation axis.

To follow the two pulse-block unitaries under the Fibonacci recursion, define
\begin{equation}
 x_n
 =
 \frac12\operatorname{Tr}
 P_s[\sigma^n(\mathsf A)],
 \qquad
 y_n
 =
 \frac12\operatorname{Tr}
 P_s[\sigma^n(\mathsf B)],
 \qquad
 z_n
 =
 \frac12\operatorname{Tr}
 \left\{
 P_s[\sigma^n(\mathsf A)]
 P_s[\sigma^n(\mathsf B)]
 \right\}.
 \label{eq:sm-fib-trace-coordinates}
\end{equation}
The first two quantities determine the rotation angles of the two
pulse-block unitaries.  Together with these two angles, $z_n$ determines the
angle between their rotation axes.  Thus $(x_n,y_n,z_n)$ contains the
information about the pair that is unchanged by rotating both axes together.

Using Eq.~\eqref{eq:sm-fib-product-recursion} and the Cayley--Hamilton
relation for $2\times2$ matrices gives
\begin{equation}
 (x_n,y_n,z_n)
 \longrightarrow
 \left(
 y_n,\,
 z_n,\,
 2y_nz_n-x_n
 \right).
 \label{eq:sm-fib-trace-map}
\end{equation}
We denote this recursion of the three half traces by
$\mathcal T_{\rm fib}$.

At pulse-block order $n=0$, Eq.~\eqref{eq:sm-fib-letter-unitary} gives
\begin{equation}
 x_0
 =
 \frac12\operatorname{Tr}A_s=s,
 \qquad
 y_0
 =
 \frac12\operatorname{Tr}B=\frac12,
 \qquad
 z_0
 =
 \frac12\operatorname{Tr}(A_sB)
 =
 s+\frac12.
 \label{eq:sm-fib-traces}
\end{equation}
Direct iteration of Eq.~\eqref{eq:sm-fib-trace-map} then gives the exact
identity
\begin{equation}
 \mathcal T_{\rm fib}^{12}
 \left(
 s,\frac12,s+\frac12
 \right)
 =
 \left(
 s,\frac12,s+\frac12
 \right)
 \label{eq:sm-fib-period12}
\end{equation}
for every $s$ in the physical interval $-1<s<1/2$.

Thus the three half traces return to their initial values after 12
Fibonacci recursion steps throughout the entire one-parameter control
family.  This statement concerns the three half traces, not the two
pulse-block unitaries themselves.  For a noncommuting pair of $SU(2)$
unitaries, the two individual half traces and the half trace of their
product determine the pair up to simultaneous conjugation.  Hence, after
12 recursion steps, the two pulse-block unitaries are related to the initial
pair by a common conjugation.  Such a common conjugation rotates both
rotation axes together and does not change the Hilbert--Schmidt norms of
the corresponding representation averages.

An important difference from the silver-mean construction below is that no
additional adjustment of the control family is needed to obtain this
finite-period return.  The physical Fibonacci parameter $s$ itself
parametrizes the period-12 family in
Eq.~\eqref{eq:sm-fib-period12}.  Consequently, a static displacement
$s=s_*+\delta$ from any finite-group resonance remains on the same
period-12 trace recursion.  This finite return period will be used in the
control-error analysis below.

\subsection{The four finite-group resonances}
\label{sec:fib-groups}

We now locate the values of $s$ for which the two elementary pulse unitaries
$A_s$ and $B$ generate a finite subgroup of $SU(2)$.
Equation~\eqref{eq:sm-half-trace-angle} is useful because a finite rotation
group contains only a discrete set of rotation angles and therefore only a
discrete set of possible half traces.

For the pulse family defined above,
Eq.~\eqref{eq:sm-fib-traces} gives
\begin{equation}
 \frac12\operatorname{Tr}A_s=s,
 \qquad
 \frac12\operatorname{Tr}B=\frac12,
 \qquad
 \frac12\operatorname{Tr}(A_sB)
 =
 s+\frac12.
 \label{eq:sm-fib-elementary-traces}
\end{equation}
If $A_s$ and $B$ generate a finite group, then $A_s$, $B$, and $A_sB$
all belong to that group, so their half traces must belong to its allowed
set.

For $-1<s<1/2$, the rotation axes of $A_s$ and $B$ are not parallel.
Moreover, $B$ is a rotation through $2\pi/3$, whereas $A_s$ is neither a
rotation about the same axis nor a half-turn about an axis perpendicular to
it.  This excludes the finite cyclic and binary-dihedral possibilities,
leaving the binary tetrahedral ($2T$), octahedral ($2O$), and icosahedral
($2I$) groups~\cite{GrimusLudl2012}.

Their corresponding rotation groups are the rotational symmetry groups of a
tetrahedron, octahedron, and icosahedron.  Converting their allowed rotation
angles to half traces using Eq.~\eqref{eq:sm-half-trace-angle}, and including
the two $SU(2)$ lifts $U$ and $-U$ of each spatial rotation, gives
\begin{align}
 2T:&\quad
 \left\{
 \pm1,0,\pm\frac12
 \right\},
 \nonumber\\
 2O:&\quad
 \left\{
 \pm1,0,\pm\frac12,\pm\frac1{\sqrt2}
 \right\},
 \nonumber\\
 2I:&\quad
 \left\{
 \pm1,0,\pm\frac12,
 \pm\frac{\varphi}{2},
 \pm\frac1{2\varphi}
 \right\}.
 \label{eq:sm-polyhedral-traces}
\end{align}

Requiring both
\[
 \frac12\operatorname{Tr}A_s=s
\]
and
\[
 \frac12\operatorname{Tr}(A_sB)
 =
 s+\frac12
\]
to belong to the appropriate set in
Eq.~\eqref{eq:sm-polyhedral-traces} leaves exactly
\begin{equation}
 s_*
 \in
 \left\{
 -\frac{\varphi}{2},
 -\frac12,
 0,
 \frac1{2\varphi}
 \right\},
 \label{eq:sm-four-points}
\end{equation}
where $\varphi=(1+\sqrt5)/2$ is the golden ratio defined above.
The binary-octahedral case gives no additional value of $s$.

The group assignments can be checked directly from relations among the
elementary pulse unitaries.  At $s_*=0$,
\begin{equation}
 A_0^2
 =
 B^3
 =
 (A_0B)^3
 =
 -\Id .
 \label{eq:sm-tet-presentation1}
\end{equation}
After identifying $U$ and $-U$, which represent the same rotation on the
Bloch sphere, these relations have rotation orders $2$, $3$, and $3$.
They generate the tetrahedral rotation group, so the corresponding subgroup
of $SU(2)$ is $2T$.

At $s_*=-1/2$,
\begin{equation}
 A_{-1/2}^3=\Id,
 \qquad
 B^3=-\Id,
 \qquad
 (A_{-1/2}B)^2=-\Id .
 \label{eq:sm-tet-presentation2}
\end{equation}
The corresponding rotation orders are $3$, $3$, and $2$, again giving the
tetrahedral rotation group and therefore $2T$.

At either
$s_*=-\varphi/2$ or $s_*=1/(2\varphi)$,
direct multiplication gives
\begin{equation}
 (A_{s_*}^2B)^2
 =
 B^3
 =
 (A_{s_*}^2B^2)^5
 =
 -\Id,
 \label{eq:sm-ico-presentation}
\end{equation}
together with
\begin{equation}
 A_{s_*}^5=\Id.
 \label{eq:sm-ico-A-order}
\end{equation}
After identifying $U$ and $-U$, the first three relations have orders
$2$, $3$, and $5$, which generate the icosahedral rotation group.
Furthermore, $A_{s_*}^2B$ and $B$ generate the same group as
$A_{s_*}$ and $B$: multiplying the first generator by $B^{-1}$ gives
$A_{s_*}^2$, and
\[
 A_{s_*}
 =
 (A_{s_*}^2)^3
\]
because $A_{s_*}^5=\Id$.  Thus both remaining resonance points generate
$2I$.

The four resonances are therefore
\begin{equation}
 \begin{array}{c|c}
 s_* &
 \langle A_{s_*},B\rangle
 \\ \hline
 -\varphi/2   & 2I\\
 -1/2         & 2T\\
 0            & 2T\\
 1/(2\varphi) & 2I
 \end{array}.
 \label{eq:sm-fib-group-table}
\end{equation}

\subsection{The first retained state moment}

We next determine the lowest spherical-harmonic degree that contains an
invariant component for each finite group.  Recall that $D^{(\ell)}$
denotes the spin-$\ell$ representation of $SU(2)$.  For integer $\ell$,
this is also the representation carried by the degree-$\ell$ spherical
harmonics on the Bloch sphere.  Since $-\Id$ acts trivially for integer
$\ell$, the invariant spaces of $2T$ and $2I$ can be calculated using the
corresponding tetrahedral and icosahedral rotation groups $T$ and $I$.

For a finite rotation group $H_{\rm rot}$, the number of invariant vectors
in $D^{(\ell)}$ is
\begin{equation}
 \dim\operatorname{Fix}_{D^{(\ell)}}(H_{\rm rot})
 =
 \frac{1}{|H_{\rm rot}|}
 \sum_{h\in H_{\rm rot}}
 \operatorname{Tr}D^{(\ell)}(h).
 \label{eq:sm-character-average}
\end{equation}
The group average on the right-hand side is the projector onto the invariant
subspace, so its trace gives the dimension of that subspace.  For a rotation
through an angle $\theta$,
\begin{equation}
 \operatorname{Tr}D^{(\ell)}(\theta)
 =
 \frac{\sin[(\ell+\tfrac12)\theta]}
      {\sin(\theta/2)}.
 \label{eq:sm-so3-character}
\end{equation}
Therefore, knowing how many rotations of each angle occur in a finite
rotation group determines whether an invariant harmonic exists at a given
$\ell$.

The tetrahedral rotation group contains the identity, eight rotations
through $2\pi/3$, and three rotations through $\pi$.  Hence
\begin{equation}
 \dim\operatorname{Fix}_{D^{(\ell)}}(T)
 =
 \frac1{12}
 \left[
 (2\ell+1)
 +
 8\frac{\sin[(\ell+\tfrac12)2\pi/3]}
        {\sin(\pi/3)}
 +
 3(-1)^\ell
 \right].
 \label{eq:sm-tet-character}
\end{equation}
For the first three nonconstant degrees,
\begin{equation}
 \dim\operatorname{Fix}_{D^{(1)}}(T)
 =
 \dim\operatorname{Fix}_{D^{(2)}}(T)
 =
 0,
 \qquad
 \dim\operatorname{Fix}_{D^{(3)}}(T)
 =
 1.
 \label{eq:sm-tet-first}
\end{equation}
Thus the first nonconstant $2T$-invariant harmonic occurs at $\ell=3$.

Similarly, the icosahedral rotation group contains the identity, fifteen
rotations through $\pi$, twenty rotations through $2\pi/3$, and twelve
rotations at each of the angles $2\pi/5$ and $4\pi/5$.  Therefore
\begin{align}
 \dim\operatorname{Fix}_{D^{(\ell)}}(I)
 =
 \frac1{60}\bigg[
 &(2\ell+1)
 +
 15(-1)^\ell
 \nonumber\\
 &+
 20
 \frac{\sin[(\ell+\tfrac12)2\pi/3]}
      {\sin(\pi/3)}
 \nonumber\\
 &+
 12
 \frac{\sin[(\ell+\tfrac12)2\pi/5]}
      {\sin(\pi/5)}
 \nonumber\\
 &+
 12
 \frac{\sin[(\ell+\tfrac12)4\pi/5]}
      {\sin(2\pi/5)}
 \bigg].
 \label{eq:sm-ico-character}
\end{align}
This quantity vanishes for $\ell=1,\ldots,5$ and equals one at $\ell=6$.
Thus the first nonconstant $2I$-invariant harmonic occurs at $\ell=6$.

The absence of an invariant through degree $k$ means that the corresponding
finite-group average agrees with the Haar state moments through that order.
Therefore the $2T$ resonances are Haar-like through state-moment order $k=2$
and can first retain a non-Haar component at $k=3$, whereas the $2I$
resonances are Haar-like through $k=5$ and can first retain a non-Haar
component at $k=6$~\cite{Delsarte1977,Reznick1995}.  These are the values
$k_*=3$ and $k_*=6$ shown in Fig.~1(c) of the main text.

\subsection{Decay of the lower-order sectors}
\label{sec:fib-rigidity}

The finite-group calculation above determines which representation sectors
contain a group-invariant component.  We must additionally verify that the
specific deterministic Fibonacci ordering does not preserve another
component that would be absent from a uniform average over the finite group.

For the substitution in Eq.~\eqref{eq:sm-fib-substitution}, the normalized
recursive averages satisfy
\begin{align}
 \overline S_\rho[\sigma^{n+1}(\mathsf B)]
 ={}&
 \frac{|\sigma^n(\mathsf B)|}
      {|\sigma^{n+1}(\mathsf B)|}
 \overline S_\rho[\sigma^n(\mathsf B)]
 \nonumber\\
 &+
 \frac{|\sigma^n(\mathsf A)|}
      {|\sigma^{n+1}(\mathsf B)|}
 \overline S_\rho[\sigma^n(\mathsf A)]
 \rho\!\left[
 P_{s_*}[\sigma^n(\mathsf B)]
 \right],
 \label{eq:sm-fib-average}
\end{align}
and
\begin{equation}
 \overline S_\rho[\sigma^{n+1}(\mathsf A)]
 =
 \overline S_\rho[\sigma^n(\mathsf B)].
 \label{eq:sm-fib-average-A}
\end{equation}
The unitary factor in the second term of
Eq.~\eqref{eq:sm-fib-average} accounts for the evolution accumulated during
the preceding block $\sigma^n(\mathsf B)$.

Equation~\eqref{eq:sm-variance-identity} shows that equality in the
norm-reduction bound requires the contributions entering each recursive
average to remain aligned.  Following this equality condition through three
consecutive Fibonacci recursion steps requires a nondecaying component to
be invariant under the two pulse-block unitaries
\begin{equation}
 P_{s_*}(\mathsf B\mathsf A)
 =
 A_{s_*}B
\end{equation}
and
\begin{equation}
 P_{s_*}(\mathsf B\mathsf A\mathsf B)
 =
 BA_{s_*}B.
\end{equation}
These two matrices generate the same subgroup as the elementary pulse
unitaries.  Indeed,
\begin{equation}
 B
 =
 P_{s_*}(\mathsf B\mathsf A\mathsf B)
 \left[
 P_{s_*}(\mathsf B\mathsf A)
 \right]^{-1},
 \label{eq:sm-fib-recover-B}
\end{equation}
and then
\begin{equation}
 A_{s_*}
 =
 P_{s_*}(\mathsf B\mathsf A)
 B^{-1}.
 \label{eq:sm-fib-recover-A}
\end{equation}
Therefore a component that avoids reduction through these Fibonacci
recursion steps must be invariant under the full subgroup
\begin{equation}
 H
 =
 \left\langle
 A_{s_*},B
 \right\rangle.
 \label{eq:sm-fib-block-group}
\end{equation}

At the $2T$ resonances no such invariant exists for $\ell=1,2$, whereas at
the $2I$ resonances none exists for $\ell=1,\ldots,5$.  At an exact
resonance all recursively generated pulse-block unitaries belong to the
finite group $H$, so only finitely many pairs of such unitaries can occur.
A strict norm reduction associated with a pair that reappears therefore
occurs repeatedly as the pulse-block order increases.  Consequently,
\begin{equation}
 \left\|
 \overline S_{D^{(\ell)}}[\sigma^n(a)]
 \right\|_{\rm HS}
 \longrightarrow0
 \qquad
 (n\to\infty)
 \label{eq:sm-fib-lower-decay}
\end{equation}
for $\ell=1,2$ at the $2T$ resonances and for
$\ell=1,\ldots,5$ at the $2I$ resonances, with
$a\in\{\mathsf A,\mathsf B\}$.

Thus the temporal averages of the actual Fibonacci sequence vanish in the
lower-order sectors required for Haar-like behavior.  This conclusion
concerns the deterministic temporal ordering itself and does not follow
merely from the corresponding uniform finite-group average.

\subsection{The physical detuning changes the retained component at first order}
\label{sec:fib-transverse}

We finally determine how the retained component changes when the physical
control parameter is displaced from a finite-group resonance.  Write
\begin{equation}
 \delta=s-s_*.
 \label{eq:sm-fib-detuning}
\end{equation}
At a tetrahedral resonance the retained harmonic has degree $\ell_*=3$,
whereas at an icosahedral resonance it has degree $\ell_*=6$.  Let $v_0$
denote the corresponding unique invariant vector in the
$D^{(\ell_*)}$ representation.

The retained nonconstant harmonic has no continuous rotational symmetry.
Thus no nonzero infinitesimal rotation can leave $v_0$ unchanged.

At each resonance, let $p$ be the smallest positive integer for which
\begin{equation}
 A_{s_*}^p
 =
 \pm\Id.
 \label{eq:sm-finite-order}
\end{equation}
For the four resonance points, $p$ is $2$, $3$, or $5$.  Write
\begin{equation}
 A_s
 =
 \cos\theta(s)\Id
 -
 i\sin\theta(s)\,
 \hat{\bm n}(s)\cdot\bm\sigma,
 \qquad
 \cos\theta(s)=s.
 \label{eq:sm-As-angle}
\end{equation}
Because $p\theta(s_*)$ is an integer multiple of $\pi$,
\begin{equation}
 \left.
 \frac{d}{ds}
 A_s^p
 \right|_{s=s_*}
 =
 \mp i\,p\,\theta'(s_*)\,
 \hat{\bm n}(s_*)\cdot\bm\sigma
 \neq0.
 \label{eq:sm-power-derivative}
\end{equation}
Thus the physical detuning moves
$A_s^p$ away from $\pm\Id$ already at first order in $\delta$.

Because $v_0$ is invariant under the finite group at $s=s_*$, it is in
particular invariant under $A_{s_*}$.  If the action of $A_s$ on $v_0$
were unchanged to first order in $\delta$, then the action of $A_s^p$ on
$v_0$ would also be unchanged to first
order.  Equation~\eqref{eq:sm-power-derivative} would then produce a
nonzero infinitesimal rotation leaving $v_0$ invariant, contradicting the
absence of continuous rotational symmetry.  Hence the action of
$A_s$ on $v_0$ changes at first order in $\delta$.

The 12-step return established in
Eq.~\eqref{eq:sm-fib-period12} means that the recursive trace dynamics
continues to repeat with a finite period when $s$ is displaced from $s_*$.
Within this finite sequence of recursive averaging steps, the unitary
factors are generated from $A_s$ and $B$.  The first-order change of $A_s$
therefore produces a
first-order difference between at least two contributions to the recursive
average when they act on the retained component.

The relevant difference between contributions is thus linear in $\delta$.
This is the $q=1$ case of the even-power result
Eq.~\eqref{eq:sm-even-power}, giving
\begin{equation}
 \Delta_{\ell_*}(\delta)
 =
 c_*\delta^2
 +
 O(\delta^3),
 \qquad
 c_*>0,
 \label{eq:sm-fib-quadratic}
\end{equation}
with $\ell_*=3$ at the two $2T$ resonances and $\ell_*=6$ at the two $2I$
resonances.

Combining this result with the relation between recursion order and physical
sequence length gives
\begin{equation}
 \log\tau_{\ell_*,\epsilon}(\delta)
 \propto
 |\delta|^{-2},
 \label{eq:sm-fib-length}
\end{equation}
where $\tau_{\ell_*,\epsilon}$ is the selected-moment lifetime defined above.
Equivalently,
\begin{equation}
 \tau_{\ell_*,\epsilon}(\delta)
 \sim
 \exp\!\left[
 \frac{\mathrm{const.}}{\delta^2}
 \right]
 \label{eq:sm-fib-exponential-length}
\end{equation}
up to factors that do not affect the leading dependence as
$\delta\rightarrow0$.

\section{Silver-mean realization}
\label{sec:silver}

The silver-mean drive provides a second example showing that the mechanism
described above is not specific to the Fibonacci recursion.  The symbols
$\mathsf A$ and $\mathsf B$ again denote letters, not matrices.  For the silver-mean drive we
use the substitution
\begin{equation}
 \sigma(\mathsf A)=\mathsf B,
 \qquad
 \sigma(\mathsf B)=\mathsf B\mathsf B\mathsf A.
 \label{eq:sm-silver-substitution}
\end{equation}
Hence
\begin{equation}
 \sigma^{n+1}(\mathsf A)=\sigma^n(\mathsf B),
 \qquad
 \sigma^{n+1}(\mathsf B)
 =
 \sigma^n(\mathsf B)\sigma^n(\mathsf B)\sigma^n(\mathsf A).
 \label{eq:sm-silver-block-recursion}
\end{equation}
The lengths of these pulse blocks grow exponentially with inflation factor
$1+\sqrt2$.

The silver-mean detuning considered below is a different one-parameter pulse
family from the Fibonacci detuning.  We therefore write its word-to-unitary
map as $P_\delta(x)$.  At the exact $2I$ resonance, $\delta=0$, we choose the
two one-pulse unitaries to coincide with those at the positive icosahedral
Fibonacci resonance:
\begin{equation}
 P_0(\mathsf A)=A_{1/(2\varphi)},
 \qquad
 P_0(\mathsf B)=B.
 \label{eq:sm-silver-resonant-pulses}
\end{equation}
Away from $\delta=0$, however, $P_\delta(\mathsf A)$ and $P_\delta(\mathsf B)$ are allowed
to vary along the silver-mean control family constructed below.  Thus the
letters $\mathsf A,\mathsf B$ remain fixed while the physical unitaries assigned to them
depend on the chosen control family.

Using Eq.~\eqref{eq:sm-product-convention}, the total unitaries produced by
the silver-mean pulse blocks satisfy
\begin{align}
 P_\delta[\sigma^{n+1}(\mathsf A)]
 &=
 P_\delta[\sigma^n(\mathsf B)],
 \nonumber\\
 P_\delta[\sigma^{n+1}(\mathsf B)]
 &=
 P_\delta[\sigma^n(\mathsf A)]\,
 P_\delta[\sigma^n(\mathsf B)]^2.
 \label{eq:sm-silver-product-recursion}
\end{align}
Thus the total unitaries of the two $n$th pulse blocks determine those of the
two $(n+1)$th pulse blocks.  If
$W_n=P_\delta[\sigma^n(\mathsf B)]$, then
$P_\delta[\sigma^n(\mathsf A)]=W_{n-1}$ and the second line of
Eq.~\eqref{eq:sm-silver-product-recursion} becomes the main-text unitary
recursion $W_{n+1}=W_{n-1}W_n^2$.

As shown in Eq.~\eqref{eq:sm-half-trace-angle}, the half trace of an
$SU(2)$ unitary determines its rotation angle.  To follow the two
pulse-block unitaries under the silver-mean recursion, define
\begin{equation}
 x_n=
 \frac12\operatorname{Tr}P_\delta[\sigma^n(\mathsf A)],
 \qquad
 y_n=
 \frac12\operatorname{Tr}P_\delta[\sigma^n(\mathsf B)],
 \qquad
 z_n=
 \frac12\operatorname{Tr}
 \left\{
 P_\delta[\sigma^n(\mathsf A)]
 P_\delta[\sigma^n(\mathsf B)]
 \right\}.
 \label{eq:sm-silver-traces}
\end{equation}
The first two quantities determine the rotation angles of the two
pulse-block unitaries.  Together with these angles, $z_n$ determines the
angle between their rotation axes.  Thus $(x_n,y_n,z_n)$ contains the
information about the pair that is unchanged when both rotation axes are
rotated together.

Using Eq.~\eqref{eq:sm-silver-product-recursion} and the Cayley--Hamilton
relation for $2\times2$ matrices gives
\begin{equation}
 (x_n,y_n,z_n)
 \longrightarrow
 \left(
 y_n,\,
 2y_nz_n-x_n,\,
 (4y_n^2-1)z_n-2x_ny_n
 \right).
 \label{eq:sm-silver-trace-map}
\end{equation}
We denote this recursion of the three half traces by
$\mathcal T_{\rm sm}$.

At the $2I$ resonance,
\begin{equation}
 (x_0,y_0,z_0)
 =
 \left(
 \frac{1}{2\varphi},
 \frac12,
 \frac{\varphi}{2}
 \right),
 \label{eq:sm-silver-point}
\end{equation}
where $\varphi=(1+\sqrt5)/2$ is the golden ratio defined above.  These are
the same three half traces as for
$P_0(\mathsf A)=A_{1/(2\varphi)}$ and $P_0(\mathsf B)=B$ in
Eq.~\eqref{eq:sm-silver-resonant-pulses}; hence the elementary pulse
unitaries generate $2I$.

Direct iteration of Eq.~\eqref{eq:sm-silver-trace-map} gives
\begin{equation}
 \mathcal T_{\rm sm}^{20}
 \left(
 \frac{1}{2\varphi},
 \frac12,
 \frac{\varphi}{2}
 \right)
 =
 \left(
 \frac{1}{2\varphi},
 \frac12,
 \frac{\varphi}{2}
 \right).
 \label{eq:sm-silver-period20}
\end{equation}
Thus the three half traces return to their initial values after 20
silver-mean recursion steps.  This statement concerns the three traces, not
the matrices themselves.  For a noncommuting pair of $SU(2)$ matrices, the
three traces determine the pair up to simultaneous conjugation.  Therefore,
when the trace triple repeats, the two pulse-block unitaries after 20 steps
are related to the initial pair by the same change of basis.  Such a common
conjugation does not change the Hilbert--Schmidt norm of the corresponding
temporal representation average.

To study a small displacement from the $2I$ resonance, we choose a nearby
one-parameter family for which this period-20 behavior of the three half
traces is preserved.  The values of the three half traces then change
smoothly with $\delta$, but return to their starting values after every
20 recursion steps.  This gives a controlled detuning for which the same
trace recursion is repeated from one 20-step cycle to the next.

We choose
\begin{equation}
 x_0(\delta)
 =
 \frac{1}{2\varphi}+\delta
 \label{eq:sm-silver-x}
\end{equation}
and determine $y_0(\delta)$ and $z_0(\delta)$ by requiring
\begin{equation}
 \mathcal T_{\rm sm}^{20}
 \left[
 x_0(\delta),y_0(\delta),z_0(\delta)
 \right]
 =
 \left[
 x_0(\delta),y_0(\delta),z_0(\delta)
 \right].
 \label{eq:sm-silver-branch}
\end{equation}
To verify that such a family exists, define
\begin{equation}
 F(x,y,z)
 =
 \mathcal T_{\rm sm}^{20}(x,y,z)-(x,y,z).
\end{equation}
A direct evaluation at the point
Eq.~\eqref{eq:sm-silver-point} shows that the period-20 condition is
satisfied and that locally only two of its three equations are independent.
For two independent components $F_1$ and $F_2$,
\begin{equation}
 \det
 \frac{\partial(F_1,F_2)}
      {\partial(y,z)}
 =
 -3480(1+\sqrt5)\ne0
 \label{eq:sm-silver-minor}
\end{equation}
at the resonance.  The implicit-function theorem therefore gives smooth
functions $y_0(\delta)$ and $z_0(\delta)$ near $\delta=0$.  Differentiating
the period-20 condition gives
\begin{equation}
 x_0'(0)=1,
 \qquad
 y_0'(0)=-\varphi,
 \qquad
 z_0'(0)=2.
 \label{eq:sm-silver-tangent}
\end{equation}

These three half traces can be realized by a smooth family of physical
$SU(2)$ pulse pairs.  The quantities $x_0$ and $y_0$ determine the two
rotation angles, while $z_0$ determines the angle between their rotation
axes.  At the $2I$ point the two axes are neither parallel nor antiparallel,
so sufficiently small changes of the three half traces can be realized by
small changes of the two pulse unitaries.  Thus
Eq.~\eqref{eq:sm-silver-branch} defines a physical one-parameter family
$P_\delta(\mathsf A),P_\delta(\mathsf B)$ passing through the $2I$ resonance.

We next verify that the specific silver-mean ordering makes the lower-order
representation averages vanish.  At $\delta=0$, the recursively generated
pulse-block unitaries satisfy
\begin{align}
 P_0[\sigma^{n+1}(\mathsf A)]
 &=
 P_0[\sigma^n(\mathsf B)],
 \nonumber\\
 P_0[\sigma^{n+1}(\mathsf B)]
 &=
 P_0[\sigma^n(\mathsf A)]P_0[\sigma^n(\mathsf B)]^2.
 \label{eq:sm-silver-resonant-recursion}
\end{align}
The pair at order $n+1$ generates exactly the same subgroup as the pair at
order $n$.  Indeed,
\begin{equation}
 P_0[\sigma^n(\mathsf A)]
 =
 P_0[\sigma^{n+1}(\mathsf B)]\,
 P_0[\sigma^{n+1}(\mathsf A)]^{-2},
\end{equation}
so both order-$n$ unitaries can be recovered from the order-$(n+1)$ pair.
Since the order-$0$ pair is $P_0(\mathsf A),P_0(\mathsf B)$, which generates $2I$, we have
\begin{equation}
 \left\langle
 P_0[\sigma^n(\mathsf A)],
 P_0[\sigma^n(\mathsf B)]
 \right\rangle
 =
 2I
 \qquad
 \text{for every }n.
 \label{eq:sm-silver-block-group}
\end{equation}

For the silver-mean substitution, the general recursive average
Eq.~\eqref{eq:sm-general-prefix-recursion} becomes
\begin{align}
 \overline S_\rho[\sigma^{n+1}(\mathsf B)]
 ={}&
 \frac{|\sigma^n(\mathsf B)|}{|\sigma^{n+1}(\mathsf B)|}
 \overline S_\rho[\sigma^n(\mathsf B)]
 \nonumber\\
 &+
 \frac{|\sigma^n(\mathsf B)|}{|\sigma^{n+1}(\mathsf B)|}
 \overline S_\rho[\sigma^n(\mathsf B)]
 \rho\!\left[P_0[\sigma^n(\mathsf B)]\right]
 \nonumber\\
 &+
 \frac{|\sigma^n(\mathsf A)|}{|\sigma^{n+1}(\mathsf B)|}
 \overline S_\rho[\sigma^n(\mathsf A)]
 \rho\!\left[P_0[\sigma^n(\mathsf B)]^2\right].
 \label{eq:sm-silver-average}
\end{align}
Here the first two contributions arise from the two consecutive copies of
$\sigma^n(\mathsf B)$ in Eq.~\eqref{eq:sm-silver-block-recursion}.  The second copy
begins after the first has been completed, producing the factor
$\rho[P_0[\sigma^n(\mathsf B)]]$, while the final $\sigma^n(\mathsf A)$ block begins after
both copies and therefore carries
$\rho[P_0[\sigma^n(\mathsf B)]^2]$.

The squared-difference identity
Eq.~\eqref{eq:sm-variance-identity} shows that repeated recursive averaging
can preserve a nonzero component only if the relevant contributions remain
aligned.  Applying this equality condition to successive silver-mean steps
requires such a component to be unchanged by both pulse-block unitaries in
Eq.~\eqref{eq:sm-silver-block-group}.  It must therefore be invariant under
$2I$.

Equation~\eqref{eq:sm-ico-character} shows that $2I$ has no invariant
component for $\ell=1,\ldots,5$, whereas its first nonconstant invariant
occurs at $\ell=6$.  Consequently,
\begin{equation}
 \left\|
 \overline S_{D^{(\ell)}}[\sigma^n(a)]
 \right\|_{\rm HS}
 \longrightarrow0,
 \qquad
 \ell=1,\ldots,5,
 \qquad
 n\to\infty,
 \label{eq:sm-silver-lower-decay}
\end{equation}
at the silver-mean $2I$ resonance.  Thus the specific deterministic ordering
of the silver-mean sequence, rather than only a uniform average over $2I$,
produces the vanishing lower-order temporal averages required for Haar-like
behavior.

Finally, we determine how the retained $\ell=6$ component changes under the
silver-mean detuning.  From Eq.~\eqref{eq:sm-silver-tangent},
\begin{equation}
 y_0'(0)=-\varphi\ne0.
\end{equation}
By definition,
\begin{equation}
 y_0(\delta)
 =
 \frac12\operatorname{Tr}P_\delta(\mathsf B).
\end{equation}
Writing the rotation angle of the elementary pulse unitary $P_\delta(\mathsf B)$ as
$2\theta_B(\delta)$ gives
\begin{equation}
 y_0(\delta)=\cos\theta_B(\delta).
\end{equation}
Since $y_0(0)=1/2$, one has $\theta_B(0)=\pi/3$ and hence
\begin{equation}
 P_0(\mathsf B)^3=-\Id.
\end{equation}
Because $y_0'(0)\ne0$, the angle $\theta_B(\delta)$ changes at first order
in $\delta$, so $P_\delta(\mathsf B)^3$ moves away from $-\Id$ already at first
order.

Let $v_0$ denote the unique $2I$-invariant vector in the $\ell=6$
representation.  As in the Fibonacci case, this nonconstant icosahedral
harmonic has no continuous rotational symmetry.  The first-order rotation
of $P_\delta(\mathsf B)^3$ therefore acts nontrivially on $v_0$, which implies that
the action of $P_\delta(\mathsf B)$ on $v_0$ itself changes at first order.

This first-order change appears directly in the recursive average
Eq.~\eqref{eq:sm-silver-average}: already for $n=0$, the first two
contributions differ by the relative factor $\rho[P_\delta(\mathsf B)]$.
At $\delta=0$ this factor leaves $v_0$ invariant, whereas for small nonzero
$\delta$ its action on $v_0$ differs from the identity by a nonzero term
linear in $\delta$.  Thus the relevant difference between contributions
appears at first order in the detuning.

This is the $q=1$ case of the even-power result
Eq.~\eqref{eq:sm-even-power}, giving
\begin{equation}
 \Delta_6^{\rm silver}(\delta)
 =
 c_I^{\rm silver}\delta^2+O(\delta^3),
 \qquad
 c_I^{\rm silver}>0.
 \label{eq:sm-silver-gap}
\end{equation}
Combining this result with the relation between the decay rate and physical
pulse length gives
\begin{equation}
 \log\tau_{6,\epsilon}^{\rm silver}(\delta)
 \propto
 |\delta|^{-2},
 \label{eq:sm-silver-length}
\end{equation}
where $\tau_{6,\epsilon}^{\rm silver}$ is the selected-moment lifetime defined
above.  Thus this lifetime grows as
$\exp[\mathrm{const.}/\delta^2]$ on approaching the silver-mean $2I$
resonance, with the same inverse-square dependence in the exponent as for
the Fibonacci drive.
\section{Additional experimental imperfections}
\label{sec:imperfections}

The main text treats the two control variations used directly in
Fig.~2(c,d): a static displacement $\bar\delta$ along the designed
resonance-crossing family and independent pulse-to-pulse fluctuations of that
same control.  On the $t$th application of the tunable Fibonacci pulse,
\begin{equation}
 \delta_t=\bar\delta+\epsilon_t,
\end{equation}
where $\epsilon_t$ has zero mean and root-mean-square magnitude
$\sigma_\delta$.  In the simulations, this means
$s_t=s_*+\bar\delta+\epsilon_t$ for each occurrence of $A_s$, while $B$
remains fixed.  The static detuning and the pulse-to-pulse fluctuations have
different effects: the former follows the smooth period-12 Fibonacci family,
whereas the latter changes independently from pulse to pulse and therefore
does not define a finite-period recursive trajectory.

We briefly discuss two additional experimental limitations that are not part
of the intrinsic detuning mechanism: errors in other pulse controls and
environmental decoherence.

\subsection{Errors in other pulse controls}

A physical single-qubit pulse generally depends on several calibrated
parameters, including its rotation angle and axis.  A small static error in a
parameter other than the designed detuning can be decomposed locally into a
component tangent to the intended one-parameter family and a component
transverse to it.  The tangent component simply changes the effective value of
$\bar\delta$.  A transverse component generally moves the pulse pair away from
the finite-period family used in the detuning analysis.

The finite-period returns established above make this distinction precise.
For Fibonacci, the three trace coordinates return after 12 recursion steps,
Eq.~\eqref{eq:sm-fib-period12}; for the silver-mean construction, the chosen
detuning family returns after 20 steps,
Eq.~\eqref{eq:sm-silver-branch}.  A small transverse calibration error induces a displacement of these trace
coordinates and can be followed by linearizing the corresponding 12- or
20-step return map.  If the induced displacement has amplitude $\eta$ along
an eigen-direction with return-map eigenvalue $\mu$,
then, while the linear approximation remains valid,
\begin{equation}
 \eta_m\simeq\mu^m\eta
 \label{eq:sm-calibration-growth}
\end{equation}
after $m$ return cycles.  Thus $|\mu|>1$ amplifies that calibration error,
whereas $|\mu|<1$ suppresses it.  The value of $\mu$ depends on the particular
laboratory control direction, so there is no universal additional lifetime
law analogous to the intrinsic
$\log\tau_{k_*,\epsilon}\propto|\bar\delta|^{-2}$ scaling.

Pulse-to-pulse fluctuations in other control directions are similarly
nonuniversal.  For weak zero-mean independent fluctuations, the ensemble
average has no first-order correction, so their leading effect is second
order in the fluctuation amplitudes.  Unlike fluctuations confined to the
designed parameter $s$, however, such errors can also generate lower-order
harmonic components.  Monitoring $R_\ell$ for $\ell<\ell_*$ together with
$R_{\ell_*}$ therefore provides a practical diagnostic: loss of the selected
signal without appreciable lower-order growth is consistent with the noise
model used in Fig.~2(c,d), whereas growth of lower harmonics indicates errors
outside that one-parameter model.

\subsection{Environmental decoherence}

Environmental decoherence provides a further cutoff that is independent of
the recursive mechanism.  Let $T_{\rm coh}$ denote the coherence scale
measured in number of applied pulses.  If the physical coherence time is
$t_{\rm coh}$ and pulses are separated by a characteristic time $t_{\rm p}$,
then parametrically
\begin{equation}
 T_{\rm coh}\sim\frac{t_{\rm coh}}{t_{\rm p}}.
\end{equation}
The selected statistical component cannot be observed for pulse numbers much
larger than $T_{\rm coh}$ even when the ideal coherent dynamics predicts a
much longer lifetime.  Standard dynamical-decoupling methods can increase
$T_{\rm coh}$ when the environmental noise spectrum permits
~\cite{ViolaLloyd1998,Uhrig2007,Cywinski2008}; this changes the experimental
cutoff, not the coherent detuning mechanism derived above.

It is therefore useful experimentally to regard the observable window as
bounded by the intrinsic selected-moment lifetime, stochastic control noise,
other calibration errors, and decoherence.  Schematically,
\begin{equation}
 T_{\rm obs}
 \lesssim
 \min\left(
 \tau_{k_*,\epsilon},
 T_{\rm cal},
 \tau_{\rm noise},
 T_{\rm coh}
 \right),
 \label{eq:sm-cutoffs}
\end{equation}
where $T_{\rm cal}$ denotes a device-dependent cutoff from other calibration
errors.  The intrinsic static-detuning law
$\log\tau_{k_*,\epsilon}\propto|\bar\delta|^{-2}$ and the independent-noise
scale $\tau_{\rm noise}\sim\sigma_\delta^{-2}$ are the two controlled
scalings emphasized in the main text; $T_{\rm cal}$ and $T_{\rm coh}$ are
extrinsic limits that should simply exceed the pulse range over which those
scalings are tested.

\end{widetext}

\end{document}